\documentclass{ieeetj}
\usepackage{cite}
\usepackage{amsmath,amssymb,amsfonts}
\usepackage{algorithmic}
\usepackage{graphicx,color}
\usepackage{textcomp}
\usepackage{xcolor}
\usepackage{hyperref}
\hypersetup{hidelinks=true}
\usepackage{comment}
\usepackage{algorithm,algorithmic}
\usepackage{tikz}
\UseRawInputEncoding

\def\BibTeX{{\rm B\kern-.05em{\sc i\kern-.025em b}\kern-.08em
    T\kern-.1667em\lower.7ex\hbox{E}\kern-.125emX}}
\AtBeginDocument{\definecolor{tmlcncolor}{cmyk}{0.93,0.59,0.15,0.02}\definecolor{NavyBlue}{RGB}{0,86,125}}

\DeclareMathOperator*{\argmin}{argmin}
\usepackage{soul,xcolor}
\usepackage{outlines}
\usepackage{subfig}
\usepackage[noabbrev]{cleveref} 
\usepackage[figurename=Figure]{caption}
\crefformat{equation}{(#2#1#3)}
\usepackage[thmmarks,amsmath]{ntheorem}
\theoremnumbering{arabic}
\theoremstyle{break}
\theoremsymbol{}
\theoremheaderfont{\normalfont\bfseries}
\theorembodyfont{\upshape}

\theoremstyle{nonumberbreak}
\usepackage{bm}
\usepackage{sansmath}
\usepackage[fleqn]{nccmath}
\usepackage[]{glossaries}
\newacronym{2d}{2D}{two-dimensional}
\newacronym{3d}{3D}{three-dimensional}
\newacronym{3gpp}{3GPP}{Third Generation Partnership Project}
\newacronym{5g}{5G}{Fifth Generation}
\newacronym{agc}{AGC}{automatic gain control}
\newacronym{aoa}{AOA}{angle-of-arrival}
\newacronym{awgn}{AWGN}{additive white Gaussian noise}
\newacronym{bp}{BP}{belief propagation}
\newacronym{bs}{BS}{base station}
\newacronym{cdf}{CDF}{cumulative distribution function}
\newacronym{csi}{CSI}{channel state information}
\newacronym{cr}{CR}{coverage region}
\newacronym{crs}{CRS}{cell-specific reference signal}
\newacronym{da}{DA}{data association}
\newacronym{dmc}{DMC}{dense multipath component}
\newacronym{eadf}{EADF}{effective aperture distribution functions}
\newacronym{ekf}{EKF}{extended Kalman filter}
\newacronym{fa}{FA}{false alarm}
\newacronym{fg}{FG}{factor graph}
\newacronym{gnss}{GNSS}{Global Navigational Satellite Systems}
\newacronym{gmf}{GMF}{global map feature}
\newacronym{gps}{GPS}{Global Positioning System}
\newacronym{iid}{iid}{independent and identically distributed}
\newacronym{imu}{IMU}{inertial measurement unit}
\newacronym{lhf}{LHF}{likelihood function}
\newacronym{los}{LoS}{line-of-sight}
\newacronym{lte}{LTE}{Long Term Evolution}
\newacronym{mb}{MB}{multi-Bernoulli }
\newacronym{mf}{MF}{map feature}
\newacronym{mimo}{MIMO}{Multiple-Input Multiple-Output}
\newacronym{miso}{MISO}{multiple-input single-output}
\newacronym{mmse}{MMSE}{minimum mean-square error}
\newacronym{mospa}{MOSPA}{mean \gls{ospa}}
\newacronym{mp}{MP}{multipath-based}
\newacronym{mpc}{MPC}{multipath component}
\newacronym{mpslam}{MP-SLAM}{multipath-based simultaneous localization and mapping}
\newacronym{nlos}{NLoS}{non-line-of-sight}
\newacronym{olos}{OLoS}{obstructed-line-of-sight}
\newacronym{ospa}{OSPA}{optimal subpattern assignment}
\newacronym{pa}{PA}{physical anchor}
\newacronym{pdf}{PDF}{probability density function}
\newacronym{pf}{PF}{potential feature}
\newacronym{phd}{PHD}{probability hypothesis density}
\newacronym{pmf}{PMF}{probability mass function}
\newacronym{ppp}{PPP}{Poisson point process}
\newacronym{pr}{PR}{physical reflector}
\newacronym{rf}{RF}{radio frequency}
\newacronym{rfs}{RFS}{random finite set}
\newacronym{rms}{RMS}{root mean square}
\newacronym{rmse}{RMSE}{root mean square error}
\newacronym{roi}{RoI}{region-of-interest}
\newacronym{rs}{RS}{reference signal}
\newacronym{rss}{RSS}{Received Signal Strength}
\newacronym{rx}{Rx}{receiver}
\newacronym{sinr}{SINR}{signal-to-interference-plus-noise ratio}
\newacronym{simo}{SIMO}{single-input multiple-output}
\newacronym{siso}{SISO}{single-input single-output}
\newacronym{slam}{SLAM}{simultaneous localization and mapping}
\newacronym{snr}{SNR}{signal-to-noise ratio}
\newacronym{spa}{SPA}{sum-product algorithm}
\newacronym{std}{STD}{standard deviation}
\newacronym{suca}{SUCA}{Stacked Uniform Circular Array}
\newacronym{tdoa}{TDOA}{time-difference-of-arrival}
\newacronym{toa}{TOA}{time-of-arrival}
\newacronym{tx}{Tx}{transmitter}
\newacronym{ue}{UE}{User Equipment}
\newacronym{usrp}{USRP}{universal software radio peripheral}
\newacronym{va}{VA}{virtual anchor}

\usepackage{multicol}
\usepackage{siunitx}
\usepackage{pgfplots}
\pgfplotsset{compat=1.11}
\usepackage{tikzscale}
\usetikzlibrary{spy,backgrounds}
\usetikzlibrary{fit}
\usetikzlibrary{arrows}
\usetikzlibrary{calc}
\usepgfplotslibrary{fillbetween}

\newcommand{\V}[1]{\bm{#1}}
\newcommand{\RV}[1]{\bm{\mathsf{#1}}}
\newcommand{\rv}[1]{\mathsf{#1}}

\newcommand{\ist}{\hspace*{.3mm}}
\newcommand{\rmv}{\hspace*{-.3mm}}
\newcommand{\iist}{\hspace*{1mm}}

\newcommand{\nn}{\nonumber}
\newcommand{\cmt}[1]{}

\providecommand{\norm}[1]{\lVert#1\rVert}

\renewcommand\IEEEkeywordsname{Index Terms}

\makeatletter
\def\OJlogo{} 
\def\seclogo{}
\makeatother

\def\authorrefmark#1{\ensuremath{^{\textbf{#1}}}}

\makeatletter
\def\@revised{}
\makeatother

\begin{document}

\begin{comment}
\receiveddate{XX Month, XXXX}
\reviseddate{XX Month, XXXX}
\accepteddate{XX Month, XXXX}
\publisheddate{XX Month, XXXX}
\currentdate{XX Month, XXXX}
\doiinfo{XXXX.2022.1234567}
\end{comment}

\markboth{Robust Localization in Modern Cellular Networks using Global Map Features}{Chen {et al.}}
\title{Robust Localization in Modern Cellular Networks using Global Map Features}

\author{Junshi Chen\authorrefmark{1}, ~\IEEEmembership{Student Member,~IEEE}, Xuhong Li\authorrefmark{1,2}, ~\IEEEmembership{Member,~IEEE}, \\ Russ Whiton\authorrefmark{3},~\IEEEmembership{Member,~IEEE}, Erik Leitinger\authorrefmark{4}, ~\IEEEmembership{Member, ~IEEE}\\ and Fredrik~Tufvesson\authorrefmark{1}, ~\IEEEmembership{Fellow,~IEEE}}
\affil{Department of Electrical and Information Technology, Lund University, 221 00 Lund, Sweden}
\affil{Department of Electrical and Computer Engineering, University of California San Diego, USA}
\affil{European Space Agency, Keplerlaan 1, NL-2200 AG Noordwijk, The Netherlands}
\affil{Institute of Communication Networks and Satellite Communications, Graz University of Technology, 8010 Graz, Austria }
\corresp{Corresponding author: Junshi Chen (email: junshi.chen@eit.lth.se).}
\authornote{This work was supported by the Swedish Innovation Agency VINNOVA through the MIMO-PAD Project (Reference number 2018-05000), in part by The Knut and Alice Wallenberg Foundation, in part by the Ericssons Research Foundation, and in part by the Strategic Research Area Excellence Center at Link\"oping--Lund in Information Technology (ELLIIT).
}

\begin{abstract}
Radio frequency (RF) signal-based localization using modern cellular networks has emerged as a promising solution to accurately locate objects in challenging environments. One of the most promising solutions for situations involving \gls{olos} and multipath propagation is multipath-based simultaneous localization and mapping (MP-SLAM) that employs map features (MFs), such as virtual anchors. This paper presents an extended MP-SLAM method that is augmented with a global map feature (GMF) repository. This repository stores consistent MFs of high quality that are collected during prior traversals. We integrate these GMFs back into the MP-SLAM framework via a probability hypothesis density (PHD) filter, which propagates GMF intensity functions over time. Extensive simulations, together with a challenging real-world experiment using LTE RF signals in a dense urban scenario with severe multipath propagation and inter-cell interference, demonstrate that our framework achieves robust and accurate localization, thereby showcasing its effectiveness in realistic modern cellular networks such as 5G or future 6G networks. It outperforms conventional proprioceptive sensor-based localization and conventional MP-SLAM methods, and achieves reliable localization even under adverse signal conditions.
\vspace{-2mm}
\end{abstract}


\maketitle
\vspace*{-15mm}

\section{INTRODUCTION}\label{sec:intro}
High-accuracy localization is an essential component for numerous modern applications, including autonomous navigation and augmented reality. \gls{gnss} serve as a primary enabler, offering satisfactory accuracy and global coverage \cite{GrovesPrinciples, braasch2023fundamentals}. However, \gls{gnss} reception is often severely degraded in indoor and urban environments, which are characterized by \gls{olos} conditions and strong multipath propagation. Localization based on cellular networks presents a promising complement to \gls{gnss}, benefiting from their dense coverage in these challenging scenarios. However, traditional triangulation and trilateration based methods are limited in such environments, as multipath effects introduce systematic impairments in both \gls{toa} and \gls{aoa} estimates of signal paths. In order to handle the challenge, the \gls{mpc} signals are utilized to increase the robustness and improve the performance of localization. Some \gls{mpc} signals interacting with the surrounding environment are reliable and stable enough to be incorporated into the framework of \gls{slam} \cite{Durrant2006slam,thrun2005pr} as \glspl{mf}. This method is termed \gls{mpslam} that jointly estimates the agent state and the unknown and time-varying number of \gls{mf} states \cite{ChSLAMOrig,Erik2019SLAM_TWC,Kim2022pmbmslam, LeiVenTeaMey:TSP2023, Kim2024TSP, GeKalXiaGarAngKimTalValWymSev:TSP2025}. The increased signal bandwidth and array aperture of cellular systems provide high temporal and spatial resolution \glspl{mpc} and thus ensure the performance of \gls{mpslam}.

\subsection{State-of-the-Art Methods}
Existing \gls{mpslam} methods either use \glspl{mpc} estimates, e.g., \glspl{toa}, \glspl{aoa}, and complex amplitudes, extracted from \gls{rf} signals as measurements \cite{ChSLAMOrig, Erik2019SLAM_TWC, Kim2022pmbmslam,LeiVenTeaMey:TSP2023, LiLeiCaiTuf:ICC2024}, or directly use \gls{rf} signals as measurements \cite{LiangTSP2025}. \gls{mpslam} methods utilizing the \glspl{mf} are categorized as a feature-based method. The \gls{va} is a widely used \gls{mf} type which represents a mirror image of \glspl{pa} (e.g., \glspl{bs}) w.r.t. a flat surface (i.e., \gls{pr}) and models signal specular reflection.\footnote{Note that other \gls{mf} types such as point scatterers \cite{ChSLAMOrig, Kim2022pmbmslam, LiLeiCaiTuf:ICC2024}, rough surfaces \cite{YuGlobecom2020, WieVenWilLei:JAIF2023, WieVenWilWitLei:Fusion2024}, or reflective surfaces \cite{Chu2022TWC, LeiVenTeaMey:TSP2023, MVASLAM_TWC2025Arxiv} enabling data fusion across propagation paths \cite{LeiVenTeaMey:TSP2023, MVASLAM_TWC2025Arxiv} can be also considered.}

\gls{mpslam} presents a high-dimensional and nonlinear inference problem, which is further complicated by measurement impairments， such as clutter and missed detections, and measurement origin uncertainties. To address these challenges, various \gls{mpslam} approaches have been developed, for instance, the \gls{ekf}-based \gls{slam}\cite{Dissanayake2001ekfslam}, Rao-Blackwellized \gls{slam}\cite{ChSLAMOrig, thrun2002fastslam}, set-based SLAM \cite{Kim2022mapfusion,Kim2022pmbmslam,GeKalXiaGarAngKimTalValWymSev:TSP2025}, and graph-based \gls{slam} methods \cite{Loeliger2004SPM,Kschischang2001FG,Erik2019SLAM_TWC}. The graph-based approach applies the message passing rules of the \gls{spa} to a \gls{fg} representing the underlying statistical model. It shows significant advantages in providing scalable and flexible solutions to high-dimensional problems in complex environments. Incorporating amplitude statistics into \gls{mpslam} \cite{erik_icc_2019, xuhong2022twc, LiLeiCaiTuf:ICC2024} is shown to enable adaptive \gls{mf} detection, thus improving robustness and scalability in dynamic application scenarios. For cellular system-based \gls{mpslam}, spectrum reuse across neighboring \glspl{bs} introduces inter-cell interference, which complicates signal modeling and processing, and subsequently degrades localization performance \cite{Lee2013icic}. To address this problem, interference mitigation methods \cite{lte_ic, rimax_ic, Sun2019icic, Khalife2019dop, Li2020dop} have been developed to estimate and subtract interference, effectively separating co-channel signals and enabling data fusion across \glspl{bs}.

In large and complex deployment environments, such as urban areas, maintaining stable performance of \gls{mpslam} becomes even more difficult. Additional proprioceptive sensors, such as \glspl{imu} and wheel encoders, are usually integrated to provide complementary motion information and improve stability. In addition, prior map information, such as a global map, can enhance \gls{mpslam} by enabling faster and more reliable initialization and convergence \cite{Markus2020vt_map, Kim2022mapfusion, Kassas2024txinaccurate}. For feature-based \gls{slam}, a global map refers to a consistent and long-term model of the \gls{rf} propagation environment, consisting of distinct and reliably detected \glspl{mf}, i.e., \gls{gmf}, serving as a stable reference frame for localization and mapping \cite{whiton2022cellular, zak2020itsm_map}. In \cite{Kim2024TSP}, a performance gain is demonstrated by employing a Poisson point process birth model for the undetected feature state, constructed from the ground-truth \gls{mf} positions perturbed with additive Gaussian noise. Another advantage of using \gls{gmf} is the mitigation of long-term drifts in proprioceptive sensors caused by the lack of external correction signals \cite{Martin2014insrss}.

Despite the advantages of adopting \glspl{gmf}, challenges also arise due to the dynamic nature of radio environments. These dynamic variations not only introduce fluctuations to \gls{mpc} parameters, such as \glspl{toa}, \glspl{aoa}, and \glspl{snr} \cite{molisch2022wireless, Flordelis2020cost2100}, but also result in missed detections and false alarms in complex scenarios, and consequently complicate the modeling of \glspl{gmf}. Moreover, \glspl{mpc} may be perceived differently due to variations in agent positions, receiver \gls{rf} front-end characteristics, and baseband signal processing algorithms. Therefore, it is crucial to develop probabilistic modeling of \glspl{gmf} to capture uncertainties from empirical measurements. It provides a more realistic representation of the radio environment, enables integration into Bayesian methods, and improves statistical modeling. However, this modeling approach has not been widely explored in existing \gls{mpslam} methods.

\subsection{Contributions}

In this work, we present a \gls{mpslam} method extended with a \gls{gmf} repository using the \gls{mpc} parameters extracted from cellular systems as measurements. The repository is constructed and subsequently reapplied in future localization loops. We employ a \gls{phd} filter \cite{mahler_smmif_2007} to propagate the intensity functions of the \glspl{gmf} over time and handle their uncertainties caused by wireless channel fading efficiently. In \cite{Jason2012Fusion}, the \gls{phd} filter was exploited to recycle tracks with low existence probabilities to reduce complexity; In contrast, our approach focuses on representing detected \glspl{mf} with high existence probabilities with the \gls{phd} filter, and these features are integrated back into the \gls{slam} framework. Since the \gls{phd} filter provides an informative prior \gls{pdf} for the new \glspl{mf} \cite{erik_icc_2019, Kim2024TSP}, they can be exploited to update both the states of the agent and the \glspl{mf}.

\begin{figure}[t!] 
\centerline{\includegraphics[width=\columnwidth]{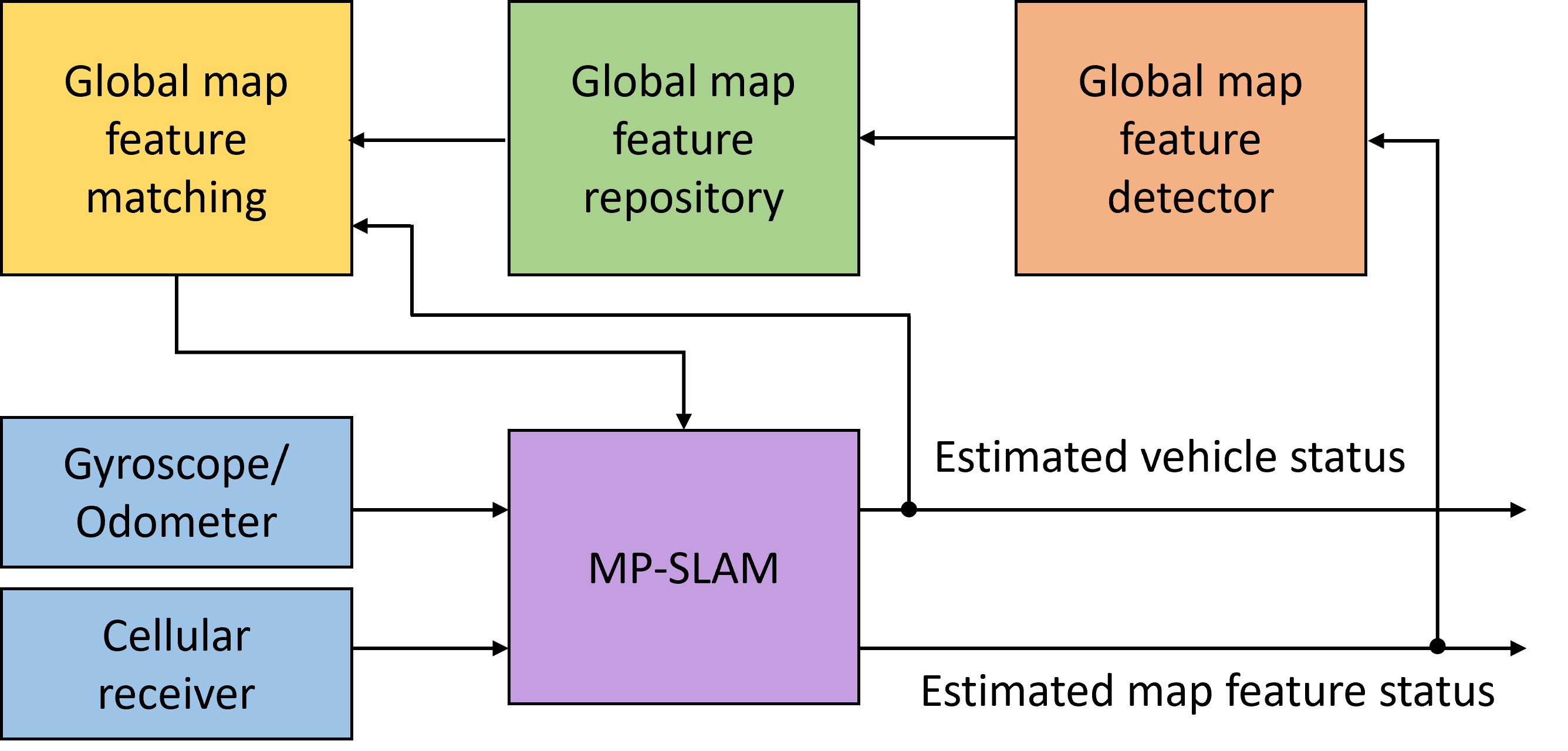}}
\caption{The system diagram of the proposed framework.} 
\label{fig:system_diagram}
\vspace{-6mm}
\end{figure}

A high-level system diagram of the proposed framework is summarized in \Cref{fig:system_diagram}. The system obtains angular rate information from a gyroscope, speed information from a wheel odometry, and \gls{mpc} information from a cellular receiver. This information is used by the \gls{mpslam} system to jointly estimate the states of the agent and \glspl{mf}. The estimated states of \glspl{mf} include their position means and variances, component \gls{snr} means and variances (where positions and component \glspl{snr} are approximated as Gaussian distributions) observed by an agent at various positions. Subsequently, these estimated states are fed into the \gls{gmf} detector to determine if a \gls{mf} qualifies for inclusion in the \gls{gmf} repository. When a closed loop is detected through \gls{gmf} matching, the valid \glspl{gmf} from the repository are fed back to \gls{mpslam} to improve the overall performance.

The main contributions of this paper are as follows.
\begin{outline}
    \1 We present a \gls{mpslam} method which exploites \gls{gmf} information from prior traversals and integrates \gls{imu} and wheel odometry information. 
    \1 We derive the \gls{gmf} implementation that includes repository establishment with detected \glspl{gmf}, identifying loop and reapplying \glspl{gmf} using \gls{phd} filters to update the states of agent and \glspl{mf}. 
    \1 We validate the proposed method using both synthetic measurements and real LTE \gls{rf} measurements collected in a challenging urban environment, demonstrating performance improvements using \glspl{gmf}.
\end{outline}
This paper focuses primarily on the application of the \gls{mpslam} method for high accuracy localization in cellular systems. The presented method is based on existing state-of-the-art methods \cite{erik_icc_2019, Erik2019SLAM_TWC}, with enhancements through \gls{gmf} features and additional sensor information.

\textit{Notations}: Column vectors and matrices are denoted as lowercase and uppercase bold letters. $f(\V{x})$ denotes the \gls{pdf} or \gls{pmf} of continuous or discrete random vector. Matrix transpose and Hermitian transpose are denoted as $(\cdot)^{\mathrm{T}}$ and $(\cdot)^{\text{H}}$, and $\mathrm{vec}(\cdot)$ means vectorization of a matrix. The operators $\odot$ and $\otimes$ denote the Hadamard product and Kronecker product. $ \norm{\cdot} $ is the Euclidean norm, and $ \vert\cdot\vert $ represents the cardinality of a set. The four-quadrant inverse tangent is denoted as $\text{atan2}(y, x)$ and inverse sine as $\text{asin}(\sqrt{(x^2+y^2)}, y)$. $\mathbf{I}_{[\cdot]}$ is an identity matrix of dimension given in the subscript, and $\V{1}_{[\cdot]}$ denotes an all-one vector. Furthermore, $ \bar{1}(a) $ denotes the function of the event $ a = 0 $ (i.e., $ \bar{1}(a) = 1 $ if $ a = 0 $, and 0 otherwise).

\section{Radio Signal Model and Channel Estimation}\label{sec:signalmodel}
We consider the radio system in a \gls{3d} scenario with horizontal and vertical signal propagation. At each discrete time $ n $, $J$ \glspl{pa} with known and fixed positions $\V{p}_{\mathrm{pa}}^{(j)} = [p_{\mathrm{pa},\mathrm{x}}^{(j)} \iist p_{\mathrm{pa},\mathrm{y}}^{(j)} \iist p_{\mathrm{pa},\mathrm{z}}^{(j)} ]^{\mathrm{T}}$, $j \in \{1,\dots,J\}$ transmit radio signals, and a mobile agent at an unknown and time-varying position $\V{p}_n=[p_{\mathrm{x},n}\, p_{\mathrm{y},n}\,p_{\mathrm{z},n}]^\mathrm{T} $ receives the signals. Each \gls{pa} and the mobile agent constitute a SIMO system, where each \gls{pa} uses a single dual-polarized antenna, and the mobile agent with unknown and time-varying heading and elevation $\varphi_{n}$ and $\theta_{n}$ uses an antenna array made of $N_{\mathrm{a}}/2$ dual-polarized elements.\footnote{The proposed algorithm can be easily reformulated for the case where the mobile agent acts as a transmitter and the \gls{pa} acts as a receiver. The extension of the algorithm to a MISO or a MIMO system considering an antenna array at the \gls{pa} side is also straightforward.} Note that $\V{p}_n$ refers to the geometric center of the array. The emitted signals interact with the surrounding objects, leading to \glspl{mpc} received by the mobile agent. The specularly reflected \glspl{mpc} can be geometrically modeled by \glspl{va} representing the mirrored positions of the \glspl{pa} w.r.t. planar surfaces (or termed \glspl{pr}). The \glspl{pa} and \glspl{va} are collectively referred to as \glspl{mf} at initially unknown but fixed positions $ \V{p}_{l}^{(j)} =[p_{l,\mathrm{x}}^{(j)} \iist p_{l,\mathrm{y}}^{(j)} \iist  p_{l,\mathrm{z}}^{(j)}]^{\mathrm{T}}$, with $l \in \{1,\dots,L_{n}^{(j)}\}$ and $ L_{n}^{(j)} $ denoting the time-varying number of visible \glspl{mf} in dynamic scenarios. To address the variables and factors related to \gls{pa}, we define $ \V{p}_{1}^{(j)} \triangleq \V{p}_{\mathrm{pa}}^{(j)} $. Frequency synchronization and constant clock offsets $\Delta\V{\tau}_{\mathrm{o}} = [\Delta\tau_{\mathrm{o}}^{(1)} \ist \cdots \ist \Delta\tau_{\mathrm{o}}^{(J)}]^{\mathrm{T}} $ between all \glspl{pa} and the mobile agent are assumed. For the \gls{mpc} associated to the $l$th \gls{mf} $\V{p}_{l}^{(j)}$, its propagation delay $\tau_{l,n}^{(j)}$, azimuth \gls{aoa} $\varphi_{l,n}^{(j)}$, elevation \gls{aoa} $\theta_{l,n}^{(j)}$, and Doppler shift $\nu_{l,n}^{(j)}$ are given by $ \tau_{l,n}^{(j)} = \|\V{p}_{n} - \V{p}_{l}^{(j)}\|/c + \Delta\tau_{\mathrm{o}}^{(j)}  $, $ \varphi_{l,n}^{(j)} = \mathrm{atan2}\big(p_{l,{\mathrm{y}}}^{(j)} - p_{\mathrm{y},n},\ist p_{l,{\mathrm{x}}}^{(j)} - p_{\mathrm{x},n}\big) - \varphi_{n} $, $\theta_{l,n}^{(j)} = \mathrm{asin}\big(\norm{\V{p}_l^{(j)}-\V{p}_n},\ist p_{l,{\mathrm{z}}}^{(j)} - p_{\mathrm{z},n}\big) - \theta_{n} $, and $\nu_{l,n}^{(j)} = (f_{\mathrm{c}}\V{v}_{l,n}^{(j)}(\V{p}_l^{(j)}-\V{p}_n)^T)/(c\ist\norm{\V{p}_l^{(j)}-\V{p}_n})$, respectively. Here $f_{\mathrm{c}}$ is the carrier frequency, $c$ is the speed-of-light, and $\V{v}_{l,n}^{(j)}$ is the agent velocity. 

\subsection{Discrete-Frequency Signal Model}\label{subsec:freq_signal_model}
By stacking the samples from all $N_{\mathrm{a}}$ antenna array elements, the discrete-frequency signal vector $\V{y}_n^{(j)} \rmv\in\rmv \mathbb{C}^{N_{\mathrm{a}}N_{\mathrm{f}}\rmv\times\rmv 1} $ from the $j$th \gls{pa} received by the mobile agent under the far-field assumption can be expressed as
\vspace{-1mm}
\begin{align}
    \hspace{-2mm}\V{y}_{n}^{(j)} &= \sum_{l=1}^{L_{n}^{(j)}} \big({\V{B}(\V{\xi}^{(j)}_{l,n})} \V{\alpha}_{l,n}^{(j)}\big) \odot \V{x}^{(j)}_{n} \rmv+\rmv {\V{b}^{(j)}_{\mathrm{dmc},n}}\odot \V{x}^{(j)}_{n} \rmv+\rmv \V{w}_{n} 
    \label{eq:SignalModel_freqDiscrete}\\[-7mm]\nn
\end{align}
where the first term describes the sum of $L_{n}^{(j)}$ specular \glspl{mpc}, with each characterized by its state vector $\boldsymbol{\xi}^{(j)}_{l,n} \triangleq [\tau_{l,n}^{(j)} \iist \varphi_{l,n}^{(j)} \iist \theta_{l,n}^{(j)} \iist \nu_{l,n}^{(j)} ]^{\mathrm{T}}$ containing the delay, azimuth and elevation \glspl{aoa}, Doppler shift, and the complex amplitudes $ \V{\alpha}_{l,n}^{(j)} \triangleq [\alpha_{\mathrm{hh},l,n}^{(j)} \iist \alpha_{\mathrm{hv},l,n}^{(j)} \iist \alpha_{\mathrm{vh},l,n}^{(j)} \iist \alpha_{\mathrm{vv},l,n}^{(j)} ]^{\mathrm{T}}$.\footnote{The subscripts $ \{\mathrm{hh}, \mathrm{hv}, \mathrm{vh}, \mathrm{vv}\} $ denote four polarimetric transmission coefficients, e.g., $ \mathrm{hv} $ indexes the horizontal-to-vertical coefficient.} We define the matrix $ \V{B}(\V{\xi}_{l,n}^{(j)}) \triangleq \big[\V{b}_{\mathrm{hh}}(\V{\xi}_{l,n}^{(j)}) \iist  \V{b}_{\mathrm{hv}}(\V{\xi}_{l,n}^{(j)}) \iist \V{b}_{\mathrm{vh}}(\V{\xi}_{l,n}^{(j)}) \iist  \V{b}_{\mathrm{vv}}(\V{\xi}_{l,n}^{(j)})\big] \rmv\in\rmv \mathbb{C}^{N_{\mathrm{a}}N_{\mathrm{f}}\rmv\times\rmv4} $ with columns given by $\V{b}_{\mathrm{hv}}(\V{\xi}_{l,n}^{(j)}) \triangleq ( \V{b}_{\mathrm{\nu}}(\nu_{l,n}^{(j)})\rmv \rmv \otimes \rmv \rmv  \V{b}_{\mathrm{f}} (\tau_{l,n}^{(j)})) \odot \mathrm{vec} \big( {b}_{\mathrm{tx},\mathrm{h}} \V{b}_{\mathrm{rx},\mathrm{v}}^{\mathrm{T}}(\varphi_{l,n}^{(j)}, \theta_{l,n}^{(j)}) \big) \in \mathbb{C}^{N_{\mathrm{a}} N_{\mathrm{f}}\rmv\times\rmv 1}$. $\V{b}_{\mathrm{\nu}}(\nu_{l,n}^{(j)}) \rmv\in\rmv \mathbb{C}^{N_{\mathrm{a}}\rmv\times\rmv 1} $ denotes the phase rotation vector due to Doppler shift, $ \V{b}_{\mathrm{f}}(\tau_{l,n}^{(j)}) \in \mathbb{C}^{N_{\mathrm{f}}\rmv\times\rmv 1} $ accounts for the system response, the baseband signal spectrum, and the phase shift due to delay $ \tau_{l,n}^{(j)} $ \cite{grebien2024SBL}; the scalar $b_{\mathrm{tx},\mathrm{h}}$ and the vector $ \V{b}_{\mathrm{rx},\mathrm{v}}(\varphi_{l,n}^{(j)}, \theta_{l,n}^{(j)}) \rmv\in\rmv \mathbb{C}^{N_{\mathrm{a}}\rmv\times\rmv N_{\mathrm{f}}} $ represent the far-field complex transmit antenna response and receive array response using the \gls{eadf} \cite{rimax_richter,Xuesong_EADF2023}. The vector ${\V{b}^{(j)}_{\mathrm{dmc},n}}$ in the second term of \eqref{eq:SignalModel_freqDiscrete} refers to the \gls{dmc} incorporating \glspl{mpc} that cannot be resolved due to the finite observation aperture. Note that the vector $\V{x}^{(j)}_{n} \in \mathbb{C}^{N_{\mathrm{a}}N_{\mathrm{f}}\rmv\times\rmv 1} $ present in both the first and second terms is defined specifically for cellular systems to account for the \gls{rs} sequence, otherwise it can be considered as an all-one vector, i.e., $\V{x}^{(j)}_{n} = \V{1}_{N_{\mathrm{a}}N_{\mathrm{f}}}$. Under a narrowband assumption, the covariance matrix of \gls{dmc} is given as a Kronecker product $\mathbf{R}_{\mathrm{dmc},n} = \mathbf{I}_{N_\mathrm{a}} \otimes \mathbf{R}_{\mathrm{f},n}$, where the spatial covariance matrix is simplified to $\mathbf{I}_{N_\mathrm{a}}$ by neglecting the spatial correlation across array elements at the agent side, and the covariance matrix in the frequency domain is characterized as $\mathbf{R}_{\mathrm{f},n}$. The third term $\V{w}_{n}$ in \eqref{eq:SignalModel_freqDiscrete} represents thermal noise that is modeled as a zero-mean, complex circular symmetric Gaussian random vector with covariance matrix $\sigma^2\mathbf{I}_{N_\mathrm{a}N_\mathrm{f}}$. The covariance matrix comprising both terms is given by $\mathbf{R}_{n} = \mathbf{R}_{\mathrm{dmc},n}+\sigma^2\mathbf{I}_{N_\mathrm{a}N_\mathrm{f}}$.

The \gls{mpc}'s component \glspl{snr} are given by $\mathrm{SNR}_{l,n}^{(j)} = \V{\alpha}_{l,n}^{(j)\text{H}} \V{B}^{\text{H}}({\V{\xi}}_{l,n}^{(j)})\mathbf{R}_{n}^{-1} \V{B}({\V{\xi}}_{l,n}^{(j)})\V{\alpha}_{l,n}^{(j)}$ and the according normalized amplitudes are given by ${u}_{l,n}^{(j)} = \mathrm{SNR}_{l,n}^{(j)\, \frac{1}{2}}$ \cite{erik_icc_2019,xuhong2022twc}.

\subsection{Parametric Channel Estimation with Interference Cancellation} \label{subsec:parametric_ce}

For cellular systems, the \gls{crs} is transmitted by the \gls{bs} to enable cell identification and channel state estimation \cite{3gpp36211}. For each \gls{bs}, the \glspl{crs} $\mathbf{x}^{(j)}_n$ are allocated to specific resource elements within the system bandwidth given the cell-ID. If cell-IDs of adjacent \glspl{bs} are congruent modulo-$3$, their \glspl{crs} occupy the same resource elements causing colliding \glspl{crs} and inter-cell interference.

To estimate the \gls{mpc} parameters from multiple \glspl{bs}, we apply a snapshot-based parametric channel estimation and detection pipeline with inter-cell interference cancellation \cite{lte_ic, rimax_ic, grebien2024SBL,MoePerWitLei:TSP2024}. First, for each transmitting antenna port, inter-cell interference between two \glspl{bs} with colliding \glspl{crs} is iteratively canceled using the SAGE--MAP algorithm \cite{lte_ic}. After this, the RIMAX algorithm \cite{rimax_richter} is applied on the cleaned signals from all receiving antennas to jointly estimate the parameters of specular \glspl{mpc} and \gls{dmc}. At each time $n$ and for each cell $j$, the \gls{mpc} parameter estimates are stacked into the vector $\vspace{-0.8mm}\V{z}_{n}^{(j)} \rmv\triangleq\rmv \big[\V{z}_{1,n}^{(j)\mathrm{T}} \ist \cdots \ist \V{z}_{M_{n}^{(j)},n}^{(j)\mathrm{T}}\big]^{\mathrm{T}} \rmv\in\rmv \mathbb{R}^{4M_{n}^{(j)}\times 1}$, where $M_n^{(j)}$ denotes the number of \glspl{mpc}. Each entry $\V{z}_{m,n}^{(j)} \rmv\triangleq\rmv \big[{z_\mathrm{d}}_{m,n}^{(j)} \iist {z_\mathrm{\varphi}}_{m,n}^{(j)} \iist {z_\mathrm{\vartheta}}_{m,n}^{(j)} \iist {z_\mathrm{u}}_{m,n}^{(j)}\big]^{\mathrm{T}} \rmv\in\rmv \mathbb{R}^{4\times 1}$ comprises the distance ${z_\mathrm{d}}_{m,n}^{(j)}$, the azimuth \gls{aoa} ${z_\mathrm{\varphi}}_{m,n}^{(j)}$, the elevation \gls{aoa} ${z_\mathrm{\vartheta}}_{m,n}^{(j)}$, and the normalized amplitude ${z_\mathrm{u}}_{m,n}^{(j)}$. Estimates $\V{z}_{m,n}^{(j)}$ with normalized amplitude ${z_\mathrm{u}}_{m,n}^{(j)}$ exceeding the threshold $u_{\mathrm{de}}$ are utilized as the (noisy) measurements.

\section{System Model}\label{sec:systemmodel}
\subsection{Agent State and Potential Map Feature States}
\label{subsec:PSMC_States}
The agent's state at time $n$ is $\RV{{x}}_n = [\RV{{p}}_n^T \iist \RV{{v}}_n^T \iist \RV{d}_{\mathrm{o}}^T ]^T$. Here, $\RV{\V{v}}_n=[\rv{s}_n \iist \rv{\psi}_n]^T$ contains the agent's speed $\rv{s}_n$ and heading $\rv{\psi}_n$ which are available from the wheel odometry and gyroscope of the agent, respectively. The distance offsets to the $J$ \glspl{bs} caused by the clock offsets are $\RV{d}_{\mathrm{o}} = \Delta\RV{\tau}_{\mathrm{o}} c=[\rv{d}_{\mathrm{o}}^{(1)} \ist \cdots \ist \rv{d}_{\mathrm{o}}^{(J)}]^T $. 

Following \cite{Erik2019SLAM_TWC, Florian2018Proceeding}, we account for the unknown and time-varying number of \glspl{mpc} by introducing \glspl{pf} indexed by $k \in \{1\ist\cdots\ist K_{n}^{(j)}\}$. The number of \glspl{pf} $K_n^{(j)}$ is the maximum number of actual \glspl{mpc} that have generated measurements so far. The existence/nonexistence of \gls{pf} $k$ is modeled by a binary random variable $ \rv{r}_{k,n} ^{(j)}\in \mathbb{{B}} = \{0,1\} $, that is, a \gls{pf} exists if and only if $r_{k,n}^{(j)} = 1$. Then the augmented states of a \gls{pf} are given as $ \RV{y}_{k,n}^{(j)} \triangleq [\RV{q}_{k,n}^{(j)\mathrm{T}} \iist \rv{r}_{k,n}^{(j)}]^{\mathrm{T}} \in \mathbb{R}^{4\times1} \times \mathbb{{B}}$, with $\RV{q}_{k,n}^{(j)} = [\RV{p}_{k}^{(j)\mathrm{T}} \iist \rv{u}_{k,n}^{(j)}]^\mathrm{T}$ including both the \gls{3d} position and the normalized amplitude. 

Formally, \gls{pf} $k$ is also considered if it is nonexistent, i.e., $r_{k,n}^{(j)} \rmv=\rmv 0$. The states $\RV{\V{q}}_{k,n}^{(j)}$ of the nonexistent \glspl{pf} are irrelevant and will not affect the \gls{pf} detection and state estimation. Therefore, the \glspl{pdf} of the nonexistent \gls{pf} states are defined as $ f(\V{q}_{k,n}^{(j)},r_{k,n}^{(j)}\rmv\rmv=\rmv\rmv 0) \rmv\rmv=\rmv\rmv f_{k,n}^{(j)} f_{\mathrm{D}}(\V{q}_{k,n}^{(j)}) $, where $ f_{\mathrm{D}}(\V{q}_{k,n}^{(j)}) $ is an arbitrary ``dummy \gls{pdf}'' and $ f_{k,n}^{(j)} \in [0,1]$ is a constant representing the probability of nonexistence \cite{Erik2019SLAM_TWC, Florian2018Proceeding, Florian2017TSP}.

\subsection{State-Transition Model}\label{sec:stm}

For each \gls{pf} with state $\RV{y}_{k,n-1}^{(j)} $ for $k \rmv\in\rmv \{1\ist \cdots\ist K_{n-1}^{(j)}\}$ at time $n-1$, there is one ``legacy'' \gls{pf} with state $ \underline{\RV{y}}_{k,n}^{(j)} \rmv\rmv\triangleq\rmv\rmv [\underline{\RV{q}}_{k,n}^{(j)\mathrm{T}} \iist \underline{\rv{r}}_{k,n}^{(j)}]^{\mathrm{T}} $ for $k \rmv\rmv\in\rmv\rmv \{1\ist \cdots\ist K_{n-1}^{(j)}\}$ at time $n$, and $\underline{\RV{q}}_{k,n}^{(j)} = [\underline{\RV{p}}_{k}^{(j)\mathrm{T}} \iist \underline{\rv{u}}_{k,n}^{(j)}]^\mathrm{T}$. The \gls{pf} states and the agent states are assumed to evolve independently, then we get
\vspace{-1mm}
\begin{align}
&f(\V{x}_n, \underline{\V{y}}_{n}|\V{x}_{n-1},\V{y}_{n-1})= f(\V{x}_n |\V{x}_{n-1})f( \underline{\V{y}}_{n}|\V{y}_{n-1}) \notag \\[-1mm]
&\hspace{10mm}= f(\V{x}_n |\V{x}_{n-1})\prod_{j=1}^{J}\prod_{k=1}^{K_{n-1}^{(j)}} f(\underline{\V{y}}_{k,n}^{(j)}|\V{y}_{k,n-1}^{(j)})
\label{eq:StateTransPDF}\\[-7mm]\nn
\end{align}
where $f(\V{x}_n |\V{x}_{n-1})$ is the agent state-transition \gls{pdf} and 
\begin{align}\label{eq:StateTransPDF_pf}
f(\underline{\V{y}}_{k,n}^{(j)}|\V{y}_{k,n-1}^{(j)})\rmv\rmv=\rmv\rmv f(\underline{\V{q}}_{k,n}^{(j)}, \underline{r}_{k,n}^{(j)}|\V{q}_{k,n-1}^{(j)}, r_{k,n-1}^{(j)})  
\end{align}
is that of the \gls{pf}. If a \gls{pf} did not exist at time $ n\rmv\rmv-\rmv\rmv1 $, i.e., $ r_{k,n-1}^{(j)}\rmv\rmv=\rmv\rmv 0 $, it cannot exist at time $ n $ as a legacy \gls{pf}, i.e.,  
\vspace{-1mm}
\begin{align}
	f(\underline{\V{q}}_{k,n}^{(j)}, \underline{r}_{k,n}^{(j)}|\V{q}_{k,n-1}^{(j)}, 0) =
	\begin{cases}
		f_{\mathrm{D}}(\underline{\V{q}}_{k,n}^{(j)}), &\underline{r}_{k,n}^{(j)}= 0\\
	0, 		      
        &\underline{r}_{k,n}^{(j)}= 1 \, .
	\end{cases}
	\label{eq:ST_pdf1}
\end{align}
On the other hand, if a \gls{pf} existed at time $ n-1 $ ($ r_{k,n-1}^{(j)} = 1 $), it either dies ($ \underline{r}_{k,n}^{(j)} = 0 $) or survives ($ \underline{r}_{k,n}^{(j)} = 1 $) with the survival probability of $P_{\mathrm{s}}$ at time $n$. If it survives, the state $ \underline{\RV{q}}_{k,n}^{(j)} $ is distributed according to the state-transition \gls{pdf} $ f(\underline{\V{q}}_{k,n}^{(j)}|\V{q}_{k,n-1}^{(j)}) $, thus, 
\vspace{-1mm}
\begin{fleqn}[0cm]
\begin{align}
f(\underline{\V{q}}_{k,n}^{(j)}, \underline{r}_{k,n}^{(j)}|\V{q}_{k,n-1}^{(j)},\! 1)\! =\!\! 
    \begin{cases}
    (1-P_{\mathrm{s}})f_{\mathrm{D}}(\underline{\V{q}}_{k,n}^{(j)}), \!\!\!\!\!       			&\underline{r}_{k,n}^{(j)}\!=\!0\\		P_{\mathrm{s}}f(\underline{\V{q}}_{k,n}^{(j)}|\V{q}_{k,n-1}^{(j)}),\!\!\!\!\!	&\underline{r}_{k,n}^{(j)}\! = \!1.  
    \end{cases}
    \label{eq:ST_pdf2}
\end{align} 
\end{fleqn}
We also define the state vector for all times up to $n$ of legacy \glspl{pf} as $ \underline{\RV{y}}_{1:n} \triangleq [\underline{\RV{y}}_{1}^{\mathrm{T}} \ist \cdots \ist  \underline{\RV{y}}_{n}^{\mathrm{T}} ]^{\mathrm{T}} $. 

\subsection{Measurements and Newly detected \glspl{pf}}\label{sec:NEWPSMC}

\gls{pf}-oriented measurements are models by the individual \glspl{lhf} $f(\V{z}_{m,n}^{(j)}\big | \V{x}_n, \V{q}_{k,n}^{(j)})$. If \gls{pf} $k$ exists ($r_{k,n}^{(j)} = 1$) it generates a \gls{pf}-oriented measurements $\V{z}^{(j)}_n$ with detection probability $P_{\text{d}}({u}_{k,n}^{(j)})$. A measurement $\V{z}_{m,n}^{(j)}$ may also not originate from any \gls{pf}.
This type of measurement is referred to as a false alarm and is modeled as a Poisson point process with mean $\mu_\text{fa}^{(j)}$ and \glspl{pdf} $f_{\text{fa}}(\V{z}_{m,n}^{(j)})$. Details are provided in Appendix \ref{sec:appdx_a}.

Newly detected \glspl{pf}, i.e., \glspl{pf} that generated a measurement for the first time, are modeled by a Poisson point process with mean $\mu^{(j)}_{\mathrm{u},n}$ and \gls{pdf} $f_{\mathrm{u},n}\big(\overline{\V{q}}^{(j)}_{m,n}\big)$ (details see Section~\Cref{sec:pf_gram}) \cite{Erik2019SLAM_TWC,Florian2018Proceeding}. Newly detected \glspl{pf} are represented by new \gls{pf} states $ \overline{\RV{y}}_{m,n}^{(j)} \triangleq [\overline{\RV{q}}_{m,n}^{(j)\mathrm{T}}\iist \overline{\rv{r}}_{m,n}^{(j)}]^{\mathrm{T}} $, $ m \in \{1\ist\cdots\ist M_n^{(j)}\} $, and $\overline{\RV{q}}_{m,n}^{(j)} = [\overline{\RV{p}}_{m}^{(j)\mathrm{T}} \iist \overline{\rv{u}}_{m,n}^{(j)}]^\mathrm{T}$. Each new \gls{pf} $ \overline{\RV{y}}_{m,n}^{(j)} $ corresponds to a measurement $\V{z}_{m,n}^{(j)}$, therefore the new \glspl{pf} number at time $n$ equals the measurement number $M_{n}^{(j)}$. If a newly detected \gls{pf} generates the measurement $\V{z}_{m,n}^{(j)}$, then $\overline{r}_{m,n}^{(j)} = 1$, otherwise $\overline{r}_{m,n}^{(j)} = 0$. The state vector of all new \glspl{pf} at time $n$ is given by $ \overline{\RV{y}}_{n}^{(j)} \triangleq [\overline{\RV{y}}_{1,n}^{(j)\mathrm{T}} \ist\cdots\ist \overline{\RV{y}}_{{M}_{n}^{(j)},n}^{(j)\mathrm{T}} ]^{\mathrm{T}} $ and the state vector for all times up to $n$ by $ \overline{\RV{y}}_{1:n}^{(j)} \triangleq [\overline{\RV{y}}_{1}^{(j)\mathrm{T}} \ist\cdots\ist \overline{\RV{y}}_{n}^{(j)\mathrm{T}} ]^{\mathrm{T}} $. The new \glspl{pf} at time $n$ become legacy \glspl{pf} at time $n+1$, and the number of legacy \glspl{pf} is then updated as $K_{n}^{(j)} = K_{n-1}^{(j)} + M_{n}^{(j)}$ accordingly (after a pruning operation, the number of \glspl{pf} is bounded). The vector containing all \gls{pf} states at time $n$ is given by $ \RV{y}_{n}^{(j)} \triangleq [\underline{\RV{y}}_{n}^{(j)\mathrm{T}} \iist \overline{\RV{y}}_{n}^{(j)\mathrm{T}} ]^{\mathrm{T}} $, where $\RV{y}_{k,n}^{(j)}$ with $ k \in \{1\ist\cdots\ist K_{n}^{(j)}\} $, and the state vector for all times up to $n$ is $ \RV{y}_{1:n}^{(j)} \triangleq [\RV{y}_{1}^{(j)\mathrm{T}} \ist\cdots\ist \RV{y}_{n}^{(j)\mathrm{T}} ]^{\mathrm{T}} $.
\subsection{Data Association Uncertainty}
\label{sec:DA}

Estimating multiple \gls{pf} states is challenging due to \gls{da} uncertainty. This is further complicated by false alarm measurements that do not correspond to any feature and missed detections of existing features. The associations between the measurements and the legacy \glspl{pf} are captured by the \gls{pf}-oriented association vector $ \RV{\underline{a}}_{n}^{(j)} \triangleq [\rv{\underline{a}}_{1,n}^{(j)} \ist \cdots \ist \rv{\underline{a}}^{(j)}_{\rv{K}_{n-1}^{(j)},n}]^{\mathrm{T}} $. If the legacy \gls{pf} $ k $ produces measurement $ m $, then $ \underline{a}_{k,n}^{(j)} \triangleq m \rmv\in\rmv \{1\ist\cdots\ist M_n^{(j)}\}$; otherwise, $ \underline{a}_{k,n}^{(j)} \triangleq 0 $. As shown in \cite{Erik2019SLAM_TWC,Florian2018Proceeding, WilliamsLau2014TAE}, the associations can equivalently be described by a measurement-oriented association vector $\vspace{-1mm}\RV{\overline{a}}_{n}^{(j)} \triangleq [\rv{\overline{a}}_{1,n}^{(j)} \ist \cdots \ist \rv{\overline{a}}_{\rv{M}_{n}^{(j)},n}^{(j)}]^{\mathrm{T}}$. If measurement $ m $ was generated by legacy \gls{pf} $ k $, then $ \overline{a}_{m,n}^{(j)} \triangleq k \rmv\in\rmv \{1\ist\cdots\ist K_{n-1}^{(j)}\} $; otherwise, $ \overline{a}_{m,n}^{(j)} \triangleq 0 $.
We assume that at any time $ n $, one \gls{pf} can generate at most one measurement, and one measurement can originate from at most one \gls{pf}. This is enforced by the exclusion functions $ \Psi(\underline{\V{a}}_n^{(j)},\V{\overline{a}}_n^{(j)}) = \prod_{k = 1}^{K_{n-1}^{(j)}}\prod_{m = 1}^{M_{n}^{(j)}}\psi(\underline{a}_{k,n}^{(j)},\overline{a}_{m,n}^{(j)})$. If $ \underline{a}_{k,n}^{(j)} = m $ and $ \overline{a}_{m,n}^{(j)} \neq k $ or $ \overline{a}_{m,n}^{(j)} = k $ and $ \underline{a}_{k,n}^{(j)} \neq m $, $ \psi(\underline{a}_{k,n}^{(j)},\overline{a}_{m,n}^{(j)}) = 0 $, otherwise it is equal to $ 1 $. The association vectors for all times up to $n$ are given by $\RV{\underline{a}}_{1:n}^{(j)} \triangleq [\RV{\underline{a}}_{1}^{(j)\mathrm{T}} \ist\cdots\ist \RV{\underline{a}}_{n}^{(j)\mathrm{T}} ]^{\mathrm{T}}$ and $ \RV{\overline{a}}_{1:n}^{(j)} \triangleq [\RV{\overline{a}}_{1}^{(j)\mathrm{T}} \ist\cdots\ist \RV{\overline{a}}_{n}^{(j)\mathrm{T}} ]^{\mathrm{T}} $. 

\subsection{Joint Posterior \gls{pdf}}
\label{sec:joing_ppdf}
Using the Bayes' rule and the independence assumptions related to the state-transition \glspl{pdf}, the prior \glspl{pdf} and the likelihoods, the joint posterior \gls{pdf} of $\RV{x}_{1:n}$, $\bar{\RV{y}}_{1:n}$, $\underline{\RV{y}}_{1:n}$, $\underline{\RV{a}}_{1:n}$, and $\RV{\overline{a}}_{1:n}$ given measurements $\V{z}_{1:n}$ for all times up to $n$ is obtained as 
\begin{align}
&f( \V{x}_{1:n}, \V{y}_{1:n}, \underline{\V{a}}_{1:n}, \V{\overline{a}}_{1:n} | \V{z}_{1:n} ) \nn \\[-1.3mm]
&\propto  f(\V{x}_{1}) \Bigg(\prod^{J}_{j'=1} \rmv \prod^{M^{(j')}_{1}}_{m'=1} \! h\big( \V{x}_{1}, \overline{\V{q}}^{(j')}_{m'\!,1}, \overline{r}^{(j')}_{m'\!,1}, {\overline{a}}^{(j')}_{m'\!,1}; \V{z}^{(j')}_{1} \big) \! \Bigg) \nn \\
&\hspace{4mm}\times \prod^{n}_{n'=2}  \! f(\V{x}_{n'}|\V{x}_{n'-1}) \prod^{J}_{j=1} \rmv\Psi\big(\underline{\V{a}}^{(j)}_{n'} \rmv,\V{\overline{a}}^{(j)}_{n'}\big)\nn\allowdisplaybreaks\\[0mm]
&\hspace{4mm}\times \rmv \Bigg( \prod^{K^{(j)}_{n'-1}}_{k=1}\rmv\rmv\rmv\big(\underline{\V{y}}^{(j)}_{k,n'} \big| \V{y}^{(j)}_{k,n'-1}\big) g\big( \V{x}_{n'}, \underline{\V{q}}^{(j)}_{k,n'} , \underline{r}^{(j)}_{k,n'}, \underline{a}_{k,n'}^{(j)}; \V{z}^{(j)}_{n'} \big)\rmv\Bigg) \nn \\
&\hspace{4mm}\times \prod^{M^{(j)}_{n'}}_{m=1} h\big( \V{x}_{n'}, \bar{\V{q}}^{(j)}_{m,n'} , \bar{r}^{(j)}_{m,n'}, \overline{a}^{(j)}_{m,n'}; \V{z}^{(j)}_{n'} \big)
\label{eq:joint_pdf}\\[-6mm]\nn
\end{align}
where the pseudo \gls{lhf} $g\big( \V{x}_{n}, \underline{\V{q}}^{(j)}_{k,n},  \underline{r}^{(j)}_{k,n}, \underline{a}^{(j)}_{k,n}; \V{z}^{(j)}_{n} \big)$ defined for existing $\underline{r}^{(j)}_{k,n}=1$ and nonexistent $\underline{r}^{(j)}_{k,n}=0$ legacy \glspl{pf} is given as 
\begin{align}
&g\big( \V{x}_{n}, \underline{\V{q}}^{(j)}_{k,n} , 1, \underline{a}^{(j)}_{k,n}; \V{z}^{(j)}_{n} \big)\nn \\
& \hspace{10mm}\triangleq
    \begin{cases}
        \frac{P_\mathrm{d}(u_{k,n}^{(j)}) f(\V{z}_{m,n}^{(j)}\big | \V{x}_{n}, \underline{\V{q}}_{k,n}^{(j)})}{ \mu_{\mathrm{fa}}^{(j)} f_{\mathrm{fa}}(\V{z}_{m,n}^{(j)})}, & \underline{a}_{k,n}^{(j)}\in \mathcal{M}_{n}^{(j)}\\
        1-P_{\mathrm{d}}(u_{k,n}^{(j)}), & \underline{a}_{k,n}^{(j)}=0\, 
    \end{cases}\label{eq:g1}\\
&g\big( \V{x}_{n}, \underline{\V{q}}^{(j)}_{k,n} , 0, \underline{a}^{(j)}_{k,n}; \V{z}^{(j)}_{n} \big)\rmv \triangleq \bar{1}(\underline{a}_{k,n}^{(j)})\,  \label{eq:g0}
\end{align}
respectively. The pseudo \gls{lhf} $h\big( \V{x}_{n}, \bar{\V{q}}^{(j)}_{m,n}, \bar{r}^{(j)}_{m,n}, \overline{a}^{(j)}_{m,n}; \V{z}^{(j)}_{n} \big)$ defined for existing $\bar{r}^{(j)}_{m,n}=1$ and nonexistent $\bar{r}^{(j)}_{m,n}=0$ new \glspl{pf} is given as
\begin{align}
&h\big( \V{x}_{n}, \bar{\V{q}}^{(j)}_{m,n} , 1, \overline{a}^{(j)}_{m,n}; \V{z}^{(j)}_{n} \big)\nn \\ 
& \hspace{3mm} \triangleq
    \begin{cases}
		0, &\overline{a}_{m,n}^{(j)}\in \mathcal{K}_{n}^{(j)}\\
		 \frac{\mu_{\mathrm{u},n}^{(j)}  f_{\mathrm{u},n}(\overline{\V{q}}_{m,n}^{(j)})f(\V{z}_{m,n}^{(j)}\big | \V{x}_{n}, \overline{\V{q}}_{m,n}^{(j)})}{\mu_{\mathrm{fa}}^{(j)} f_{\mathrm{fa}}(\V{z}_{m,n}^{(j)})},  & \overline{a}_{m,n}^{(j)} = 0 
    \end{cases} \label{eq:h}\\
&h\big( \V{x}_{n}, \bar{\V{q}}^{(j)}_{m,n} , 0, \overline{a}^{(j)}_{m,n}; \V{z}^{(j)}_{n} \big) \triangleq f_{\mathrm{D}} (\bar{\V{q}}_{m,n}^{(j)}) \, .\label{eq:h0}
\end{align}
The factorization of \cref{eq:joint_pdf} is represented by the \gls{fg} \cite{Loeliger2004SPM,Kschischang2001FG} illustrated in \Cref{fig:factorGraph} in Appendix~\ref{sec:appdx_b}. A detailed derivation of the joint posterior \gls{pdf} in \cref{eq:joint_pdf} can be found in \cite{Erik2019SLAM_TWC}.

\subsection{Building and Utilizing Global Map Feature}
\label{sec:pf_gram}

In \cite{Horridge2011phd,Jason2012Fusion, erik_icc_2019,Kim2024TSP}, a \gls{phd} filter \cite{mahler_smmif_2007} is introduced to propagate the intensity function of the undetected \glspl{mf}. For the $j$th \gls{pa} at time $n$, it has the intensity function of $\lambda_n^{\mathrm{u}}(\V{q}_{\cdot,n}^{(j)})=\mu_{\mathrm{u},n}^{(j)}f_{\mathrm{u},n}(\V{q}_{\cdot,n}^{(j)})$. The mean number $\mu_{\mathrm{u},n}^{(j)}$ and \gls{pdf} $f_{\mathrm{u},n}(\V{q}_{\cdot,n}^{(j)})$ are used in \Cref{eq:h}. 

At the beginning time $n=1$, the state of undetected \glspl{mf} for PA $j$ follows a Poisson \gls{rfs} with intensity function $\lambda_n^{\mathrm{u}}(\V{q}_{\cdot,1}^{(j)})$. In the absence of prior information on the spatial distribution of \glspl{mf}, $\lambda_n^{\mathrm{u}}(\V{q}_{\cdot,1}^{(j)})$ is assumed to be constant over the \gls{roi} and its integral over the whole \gls{roi} equals the expected number of \glspl{mf} within the \gls{roi}.

Over time, some \glspl{mf} emerge and then fade away due to blockage or increased distance. Because many of them come from the reflectors in the surrounding environment, such as buildings and windows, they tend to remain stable over time. Therefore, the information from these \glspl{mf} can be used to build a repository. The repository includes the positions and the normalized amplitudes of \glspl{gmf}, which are modeled as multidimensional Gaussian distributions. Specific criteria are applied to ensure the quality of \glspl{gmf}. For example, the \gls{gmf}'s normalized amplitudes should remain sufficiently high throughout their lifespans, which must also be long enough, and the \gls{gmf}'s estimated position variances should remain within an acceptable range. In addition, all the indices $n^{\prime}$ of the agent positions that are visible to the $g$-th \gls{gmf} constitute an index set $\mathrm{IS}_g$, and corresponding agent positions constitute a \gls{cr} $ \mathrm{CR}_g = \{\V{p}_{n^{\prime}}, n^{\prime}\in \mathrm{IS}_g\}$.     

With the repository containing \glspl{gmf}, the state propagation of the undetected \glspl{mf} can be represented as the intensity function propagation of the \gls{phd} filter \cite{Jason2012Fusion} as described in the following.
\subsubsection{\gls{gmf} Intensity Function Initialization}
\label{sec:GMASave} 
A straightforward approach to utilize \glspl{gmf} is to apply all of them immediately after they are added to the repository. However, this strategy will activate the \glspl{gmf} that are not likely to appear in a specific scene, thus increasing computational complexity. Here, we propose a simplified method that applies to most scenarios. When the agent revisits a previously explored scenario, it can utilize prior information stored in the \gls{gmf} repository. Upon approaching the same \gls{cr} and occupying a position $\V{p}_n$, the agent's Euclidean distances to all stored positions in the set $\mathrm{IS}_g$ are evaluated. The index of the closest entry is identified as
\vspace{-1mm}
\begin{align}\label{eq:gma_dist_criteria}
    n_{\mathrm{min}} = \argmin_{n^{\prime} \in \mathrm{IS}_g} 
    \| \V{p}_n - \V{p}_{n^{\prime}} \| \,.\\[-7mm]\notag
\end{align}
If the minimum distance satisfies $\| \V{p}_n - \V{p}_{n_{\mathrm{min}}} \| < d_\mathrm{min}$, the corresponding \gls{gmf} is reinserted into the prior distribution using the parameters associated with index $n_{\mathrm{min}}$. Specifically, the $g$th \gls{gmf} of the $j$th \gls{pa} adopts the intensity function
\[
\lambda_{\mathrm{g},n}(\V{q}_{\cdot,n}^{(j)}) = \mu_{\mathrm{g},n}^{(j)} f_{\mathrm{g},n}(\V{q}_{\cdot,n}^{(j)})
\]
where the spatial density $f_{\mathrm{g},n}(\V{q}_{\cdot,n}^{(j)})$ follows the form
\vspace{-1mm}
\begin{align}\label{eq:gma_dist}
f_{\mathrm{g},n}(\V{q}_{\cdot,n}^{(j)}) = \prod_{\mathrm{v}\in\{x,y,z,u\}} 
\frac{\exp\left(-\frac{(q_{\cdot,n,\mathrm{v}}^{(j)} - m_{g,n_{\mathrm{min}},\mathrm{v}}^{(j)})^2}{2\sigma^2_{g,n_{\mathrm{min}},\mathrm{v}}}\right)}{\sqrt{2\pi}\sigma_{g,n_{\mathrm{min}},\mathrm{v}}} \,.\\[-7mm]\notag
\end{align}
Here, $m_{g,n_{\mathrm{min}},\mathrm{v}}^{(j)}$ and $\sigma_{g,n_{\mathrm{min}},\mathrm{v}}$ denote the mean and \gls{std} of the selected \gls{gmf}, respectively. The index $\mathrm{v} \in \{x, y, z\}$ refers to spatial coordinates, while $\mathrm{v} = u$ represents the normalized amplitude.

\subsubsection{Prediction Step}\label{subsec:phd_predict}
The intensity function of the undetected \gls{mf} is predicted by
\begin{align}\label{eq:phd_predict}
\lambda_{{\mathrm{u}},n|n-1}(\V{q}_{\cdot,n}^{(j)}) &= \lambda_{b}(\V{q}_{\cdot,n}^{(j)}) + P_{{\mathrm{u}},\mathrm{s}}\int f(\V{q}_{\cdot,n}^{(j)}| \V{q}_{\cdot,n-1}^{\prime(j)})\notag \\
&\hspace{5mm}\times\lambda_{{\mathrm{u}},n-1}(\V{q}_{\cdot,n-1}^{\prime (j)})\mathrm{d}\V{q}_{\cdot,n-1}^{\prime (j)} \, \\[2mm]
\lambda_{b}(\V{q}_{\cdot,n}^{(j)}) &= \lambda_{b'}(\V{q}_{\cdot,n}^{(j)}) + \sum_{g=1}^G\lambda_{\mathrm{g},n}(\V{q}_{\cdot,n}^{(j)})\,  \notag\\[-7mm]
\end{align}
where $\lambda_{{\mathrm{u}},n-1}(\V{q}_{\cdot,n-1}^{(j)})$ is the intensity function of time $n-1$, $P_{{\mathrm{u}}, \mathrm{s}}$ is the survival probability, and $f(\V{q}_{\cdot,n}^{(j)}| \V{q}_{\cdot,n-1}^{\prime (j)})$ is the state-transition \gls{pdf} from $\V{q}_{\cdot,n-1}^{\prime (j)}$ to $\V{q}_{\cdot,n}^{(j)}$. $\lambda_b(\V{q}_{\cdot,n}^{(j)})$ consists of the birth intensity function $\lambda_{b'}(\V{q}_{\cdot,n}^{(j)})$ that models the birth of new \glspl{mf}, and the intensity function $\lambda_{\mathrm{g},n}(\V{q}_{\cdot,n}^{(j)})$ that models the recurrence of the $g$-th \gls{gmf} at time $n$, which is initialized by \cref{eq:gma_dist}. 

The \gls{pdf} $f_{{\mathrm{u}},n}(\V{q}_{\cdot,n}^{(j)})$ for the newly detected \glspl{mf} is obtained from $\lambda_{{\mathrm{u}}, n|n-1}(\V{q}_{\cdot,n}^{(j)})$ as
\vspace{-1mm}
\begin{align}\label{eq:phd_pdf}
    f_{\mathrm{u},n}(\V{q}_{\cdot,n}^{(j)}) = \frac{P_{{\mathrm{u}},\mathrm{d}}^{(j)}\lambda_{{\mathrm{u}},n|n-1}(\V{q}_{\cdot,n}^{(j)})}{\int P_{{\mathrm{u}},\mathrm{d}}^{(j)}\lambda_{{\mathrm{u}}, n | n-1}(\V{q}_{\cdot,n}^{\prime (j)})\mathrm{d}\V{q}_{\cdot,n}^{\prime (j)}}\\[-7mm] \notag
\end{align}
where $P_{{\mathrm{u}},\mathrm{d}}^{(j)}$ is the detection probability of undetected \glspl{mf}. The mean number of newly detected \glspl{mf} is given by
\begin{align}\label{eq:phd_meannum}
    \mu_{\mathrm{u},n}^{(j)} = \int  P_{\mathrm{u,d}}^{(j)}\lambda_{\mathrm{u},n|n-1}(\V{q}_{\cdot,n}^{\prime (j)})\mathrm{d}\V{q}_{\cdot,n}^{\prime (j)}.
\end{align}
\vspace{-3mm}
\subsubsection{Update Step}\label{subsec:phd_update}
Using the predicted intensity function $\lambda_{\mathrm{u},n|n-1}(\V{q}_{\cdot,n}^{(j)})$, the updated intensity function $\lambda_{\mathrm{u},n}(\V{q}_{\cdot,n}^{(j)})$ is given by
\begin{align}
    \lambda_{\mathrm{u},n}(\V{q}_{\cdot,n}^{(j)}) = (1-P_{\mathrm{u},d}^{(j)})\lambda_{\mathrm{u},n|n-1}(\V{q}_{\cdot,n}^{(j)}).
\end{align}
\subsection{Minimum Mean Square Error Estimation}
\label{sec:mmse_est}
The goal is to estimate the agent's position $\V{x}_n$, and the positions $\V{p}_{k}^{(j)}$ and amplitudes $u_{k,n}^{(j)}$ of \glspl{pf} from the measurements $\V{z}_{1:n}$, based on their marginal posterior \glspl{pdf}. More specifically, the estimates are obtained by using the \gls{mmse} estimator \cite{Kay_EstimationTheory}
\begin{align}
\label{eq:x_mmse}
\begin{split}
\hat{\V{x}}_n^{\mathrm{MMSE}} &\triangleq \int \V{x}_n f(\V{x}_n | \V{z}_{1:n})\mathrm{d} \V{x}_n \, 
\end{split}\\
\label{eq:q_mmse}
\begin{split} 
\hat{\V{p}}_{k}^{(j)\mathrm{MMSE}} &\triangleq \int \V{p}_{k}^{(j)}f(\V{q}_{k,n}^{(j)}| r_{k,n}^{(j)}=1, \V{z}_{1:n})\mathrm{d}\V{q}_{k,n}^{(j)} \, 
\end{split} \\
\label{eq:u_mmse}
\begin{split}
\hat{u}_{k,n}^{(j)\mathrm{MMSE}} &\triangleq \int u_{k,n}^{(j)}f(\V{q}_{k,n}^{(j)} |  r_{k,n}^{(j)}=1, \V{z}_{1:n})\mathrm{d}\V{q}_{k,n}^{(j)} 
\end{split}
\end{align}
with
\begin{align}
& f(\V{q}_{k,n}^{(j)}| r_{k,n}^{(j)}=1, \V{z}_{1:n}) = \frac{f(\V{q}_{k,n}^{(j)}, r_{k,n}^{(j)}=1| \V{z}_{1:n})}{p( r_{k,n}^{(j)}=1| \V{z}_{1:n})}\,  \\ 
&  p(r_{k,n}^{(j)}=1| \V{z}_{1:n}) = \int f(\V{q}_{k,n}^{(j)}, r_{k,n}^{(j)}=1| \V{z}_{1:n})\mathrm{d}\V{q}_{k,n}^{(j)} \,  .
\end{align}
A \gls{pf} is declared to exist if its existence probability is higher than a threshold $P_\mathrm{det}$, i.e., $p(r_{k,n}^{(j)}=1| \V{z}_{1:n})>P_\mathrm{det}$. The \glspl{pdf} $f(\V{x}_n | \V{z}_{1:n})$ and $f(\V{q}_{k,n}^{(j)}| r_{k,n}^{(j)}=1, \V{z}_{1:n})$ in \cref{eq:x_mmse,eq:q_mmse,eq:u_mmse} are marginal \glspl{pdf} of the joint posterior \gls{pdf} in \cref{eq:joint_pdf}. Since they cannot be obtained analytically, we calculate the beliefs $q(\V{x}_{n})$, $ \underline{q}(\underline{\V{q}}^{(j)}_{k,n}, \underline{r}^{(j)}_{k,n}) $, $\overline{q}(\bar{\V{q}}^{(j)}_{m,n}, \bar{r}^{(j)}_{m,n}) $ approximating the marginal \glspl{pdf} for the agent, the legacy and new \glspl{pf}, by using a particle-based message passing implementation on the \gls{fg} representing the proposed statistical model. The belief $q(\V{x}_{n})$ approximating $ f(\V{x}_{n}|\V{z}_{1:n} )$ is calculated as shown in Appendix \ref{sec:appdx_b}. The other messages and beliefs are calculated in line with \cite{Erik2019SLAM_TWC,erik_icc_2019}.

\section{Evaluation}
The performance of the proposed \gls{mpslam} method is validated using both synthetic and real \gls{rf} measurements. To analyze the performance gain from exploiting \glspl{gmf}, we compare the simulation results under three different setups: (i) \textit{Proprioception}: localization uses the information from the gyroscope and wheel odometry alone; (ii) \textit{\gls{slam} without \gls{gmf}}: \gls{slam} uses proprioceptive sensors and \gls{mpc} estimates, such as distances, azimuth and elevation \glspl{aoa}, and normalized amplitudes, as measurements; (iii) the proposed \textit{\gls{slam} with \gls{gmf}}: \gls{slam} as in (ii) but also exploits the \gls{gmf} information from early traversals. 

Assume that the agent moves on a 2D plane, i.e., the height $p_{z,n}$ remains constant over time, and its state-transition model is defined as \cite{GrovesPrinciples}
\begin{align}\label{eq:agent_state_transit}
    \begin{bmatrix}
    p_{x,n}\\
    p_{y,n}     
    \end{bmatrix} &=    \begin{bmatrix}
    p_{x,n-1} \\
    p_{y,n-1}     
    \end{bmatrix}
    + s_{n-1}\Delta_t  
    \begin{bmatrix}
    \cos({\psi_{n-1}})\\
    \sin({\psi_{n-1}}) 
    \end{bmatrix} \,   \\
    s_{n} & = s_{\mathrm{o},n} + w_{v,n}  \,  \\
    \psi_{n} & = \psi_{n-1} + \Delta_t \dot{\psi}_{n-1} + w_{\psi,n} 
\end{align}
where $\Delta_t$ is the duration of one snapshot, $s_{\mathrm{o},n}$ is the speed derived from wheel odometry, $\dot{\psi}_{n}$ is the heading rate at time $n$, $w_{s,n}$ and $w_{\psi,n}$ denote the noises which are zero--mean
and Gaussian with \glspl{std} of $\sigma_{s}$ and $\sigma_{\psi}$, respectively.

The following parameters are used in both synthetic and real measurement evaluation unless otherwise stated. The state-transition \glspl{pdf} of \glspl{pf}, including the normalized amplitudes, is set similarly to \cite{Erik2019SLAM_TWC, xuhong2022twc}. The survival probability $P_\mathrm{s}$ of one \gls{pf} is $0.9$. A \gls{pf} with an existence probability lower than $\num{e-3}$ is pruned. To maintain high quality \glspl{gmf}, a \gls{pf} has been detected for over $9.75$~seconds ($130$ snapshots) and has an $x$-axis \gls{std} smaller than $0.5$~meters is added to the \gls{gmf} repository. The minimum distance $d_\mathrm{min}$ is set to $5$\,meters. For the \gls{phd} filter, the new \gls{mf} birth intensity $\lambda_b$ is $\num{e-5}$, the survival probability $P_{{\mathrm{u}},\mathrm{s}}$ is $0.9$, and the detection probability $P_{{\mathrm{u}},\mathrm{d}}$ is $0.1$. The particle number is $\num{e6}$, and the simulation runs for $50$ iterations with different random seeds.

\subsection{Performance of Synthetic \gls{rf} Measurements}\label{sec:sync_data_sim}
\begin{figure}[!t]
\centering
\includegraphics{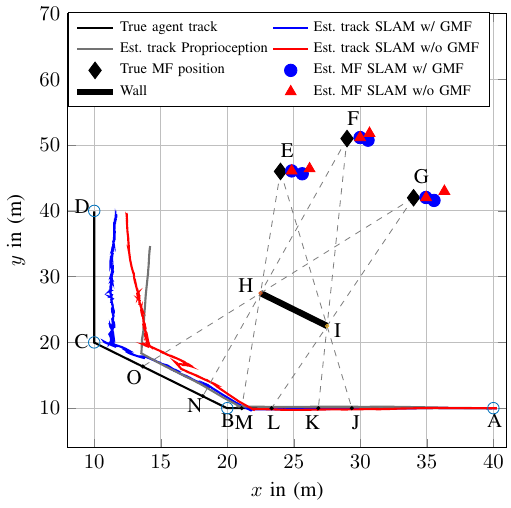}
\caption{
Geometric illustration of the synthetic simulation scenario, comparing proprioception, \gls{mpslam} without and with \glspl{gmf}. The \gls{mf} positions correspond to their final estimated states. Gray dashed lines indicate segments of the true agent trajectory where signals from each \gls{mf} are blocked by the central wall.
}
\label{fig:perf_overview}
\vspace{-5mm}
\end{figure}

\begin{figure*}[!t]
\centering
\hspace{-2mm}\begin{minipage}[b]{0.235\textwidth}
\centering
\includegraphics{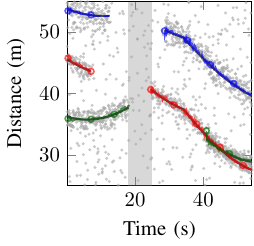}
\end{minipage}
\hfill
\begin{minipage}[b]{0.235\textwidth}
\centering
\includegraphics{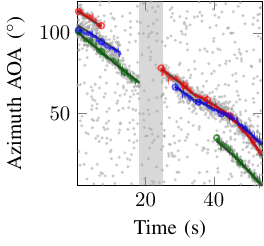}
\end{minipage}
\hfill
\begin{minipage}[b]{0.235\textwidth}
\centering
\includegraphics{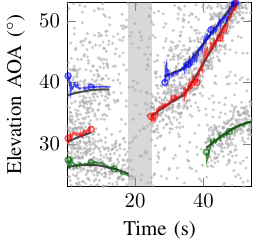}
\end{minipage}
\hfill
\begin{minipage}[b]{0.235\textwidth}
\centering
\includegraphics{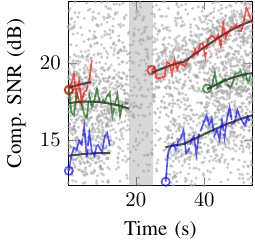}
\end{minipage}
\caption{Results of a simulation run using synthetic measurements at $\text{SNR}_{\text{1m}} = 33.5 \, \text{dB}$. The estimated parameters (dashed lines with circle markers) are compared to the ground truth (black solid lines). Each color corresponds to one \gls{mf}. Gray dots indicate the measurements, including false alarms. The shaded gray region highlights the fully blocked segment. }
\label{fig:slam_est_para}
\vspace{-5mm}
\end{figure*}

To verify the proposed algorithm, we generate synthetic \gls{mpc} measurements $\V{z}_{n}$ according to the environment shown in \Cref{fig:perf_overview}, where a SIMO system operating at center frequency $2.6$\,GHz with effective bandwidth of $18$\,MHz is used. The agent is equipped with a $128$-port \gls{suca} array. The agent moves from point A to point D via turning points B and C with the speed of $\qty{1}{m/s}$. There are $3$ \glspl{mf}, i.e., E, F, and G in the scene (E is a \gls{pa}), with \gls{3d} coordinates of $[24, 46, 23], [29, 51, 33]$ and $[34, 42, 16]$. A wall between points H and I blocks the signals from \glspl{mf} in parts of the agent's trajectory. As a result, the agent receives signals from $0$ to $3$ \glspl{mf} along the trajectory. For example, it cannot receive any signal in the segment between points L and M (the corresponding times are $18$ and $24.6$\,seconds). The agent will diverge from the track within this segment with only proprioception information and will have difficulty returning to the track without \gls{gmf} information. 

Corresponding to $3$ \glspl{mf}, $3$ \glspl{mpc} with time-varying parameters are synthesized. The amplitude of each \gls{mpc} is assumed to follow free-space path loss and is further attenuated by $3$\,dB after each reflection. To obtain the \gls{mpc}'s component \glspl{snr}, \gls{awgn} measurement noise is generated for each simulation run with variance $\sigma^2$ specified by the output \gls{snr}, i.e., $\mathrm{SNR}_{1\mathrm{m}} = 10\log10(\frac{|\alpha_{\mathrm{LOS}}|^2\|\V{s}_{\mathrm{LOS}}\| ^2}{\sigma^2} )$ including array gain and frequency sample gain. The amplitude $\alpha_{\mathrm{LOS}}$ and the signal vector $\V{s}_{\mathrm{LOS}}$ of the \gls{los} path are calculated at a distance of $1$\,meters. We perform $50$ simulation runs for each $\mathrm{SNR}_{1\mathrm{m}} \in \{33.5, 37.5, 41.5\}$\,dB. Since the agent's distance to the \gls{bs} is around $50$ meters, it has an additional $17$ dB of path loss, and the actual received \glspl{snr} for \glspl{los} are $\{16.5, 20.5, 24.5\}$\,dB, respectively. The predefined \gls{snr} threshold for adopting an \gls{mpc} estimate as a noisy measurement is $u_{\mathrm{de}} = 12$\,dB. The noise \glspl{std} of the wheel odometry $\sigma_s$ and the gyroscope $\sigma_{\psi}$ are set to $\qty{0.19}{m/s}$ and $\qty{0.18}{\degree}$, respectively. The agent is initialized with a Gaussian position offset of mean $\{0.1, \, 0.1,\, 0\}$\,meters and \gls{std} $0.2$\,meters, and a Gaussian heading offset of mean  $0.2$\,degrees and \gls{std} $0.2$\,degrees. The \gls{roi} is a spherical segment with a radius $R = 60$\,meters and a height $H=30$\,meters, yielding a volume of \gls{roi} $2\pi(R^2-H^2/3)H = \qty{6.7676e05}\,{\mathrm{m}^3}$. The false alarm measurements are modeled by a Poisson point process with a mean value of $1$. 

We first show single simulation run \gls{mmse} estimates of the agent track and the \glspl{mf} in \Cref{fig:perf_overview}. As illustrated, the localization performance is comparable across all three simulation setups initially, following the true agent trajectory accurately until the agent reaches point L, where signals from all three \glspl{mf} are blocked. Before this period, the proposed method \textit{SLAM with GMF} gradually establishes a global map using the detected \gls{mf} information. Once the \glspl{mf} reappear after point M, it leverages this map for faster feature redetection and convergence, leading to significantly improved localization performance and more accurate \gls{mf} state estimates than the reference methods. The performance from exploiting \glspl{gmf} is further demonstrated in \Cref{fig:slam_est_para}, which shows the estimated \gls{mpc} parameters in different dimensions. The distance and angular estimates are derived via geometric transformations from the agent and \gls{mf} estimates in \Cref{fig:perf_overview}. Despite the noisy measurements with false alarms and missed detections, the \glspl{mf} are accurately detected and estimated. Especially after $24.6$\,seconds, i.e., after the blockage period in the gray shaded region, the \glspl{mf} are quickly redetected with the informative prior \gls{pdf} from the \glspl{gmf}, and the estimates closely follow the ground truth.

\begin{figure}[!t]
\centering
\hspace{-2mm}\includegraphics{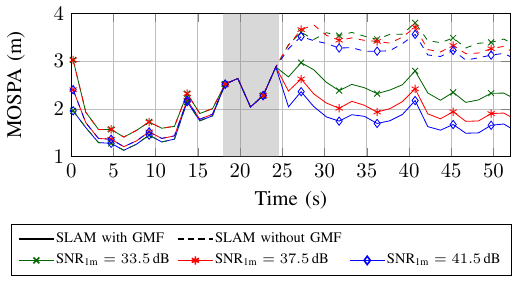}\\[0mm]
\caption{Results for synthetic measurements. The \gls{mospa} errors of the \glspl{mf} over time for \gls{slam} with and without \gls{gmf} under different \glspl{snr}, respectively.}
\label{fig:mospa}
\vspace{-10mm}
\end{figure}

In the following, we present the statistical performance evaluation of \glspl{mf} using the \gls{ospa} \cite{Vo2008OSPA} metric, which can efficiently capture the estimation errors of the \gls{mf} states to the true \gls{mf} states at each time step. We use \gls{ospa} metric order one and set the cutoff distance to $6$ meters, which denotes the weighting of an estiamted \gls{mf} that does not match with a true \gls{mf}. \Cref{fig:mospa} shows the \gls{mospa} errors of the \glspl{mf} averaging over all simulation runs for different \glspl{snr} and time steps, respectively. When $\text{SNR}_{\text{1m}}$ increases, \gls{mospa} decreases for both \glspl{slam} with and without \gls{gmf}. In addition, \textit{\gls{slam} with \gls{gmf}} significantly outperforms \textit{\gls{slam} without \gls{gmf}} after the signal blockage period around $24.6$ seconds, which is attributed to the \gls{gmf} information that greatly improves \gls{mf} redetection and estimation. 

\begin{figure}[!t]
    \centering
    \hspace{8.8mm}\subfloat{\scalebox{1}{\hspace{-10.3mm}\includegraphics{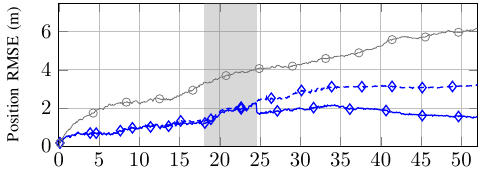}}\label{fig:syn_abs_pos_error}} \\[-2mm]
    \hspace{8.8mm}\subfloat{\scalebox{1}{\hspace{-10.3mm}\includegraphics{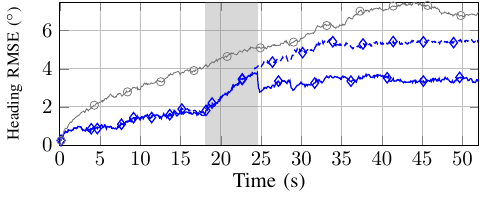}}\label{fig:syn_abs_head_error}} \\[-4mm]
    \hspace{8.5mm}\subfloat{\scalebox{1}{\hspace{-8.5mm}\includegraphics{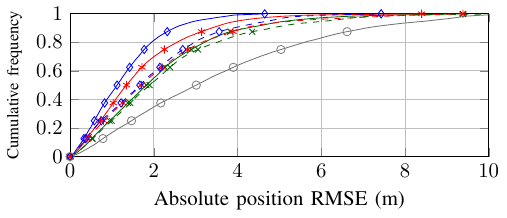}}\label{fig:cdf_pos_error}} \\[-4mm]
    \hspace{0.0mm}\subfloat{\scalebox{1}{\hspace{-0.0mm}\includegraphics{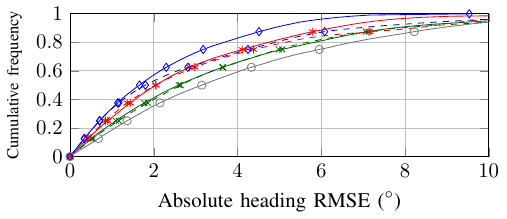}}\label{fig:cdf_heading_error}} \\[0mm]    {\hspace{-0.0mm}\includegraphics[width=\columnwidth]{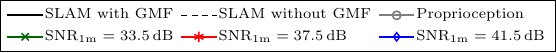}} \\[0mm]    
    \caption{Results for synthetic measurements. The \glspl{rmse} of the agent's absolute position and heading over time are shown in the first two images, for the proprioception approach, and \gls{slam} without and with \glspl{gmf}. The cumulative frequencies of \glspl{rmse} are shown in the last two images for agent position and agent heading across different settings.}\label{fig:syn_rmse_cdf}	
    \vspace{-6mm}
\end{figure}

The \glspl{rmse} of the agent's position and heading for three setups over time at $\text{SNR}_{\text{1m}}=41.5\,\text{dB}$ are presented in the first two images of \Cref{fig:syn_rmse_cdf}, respectively. \textit{Proprioception} exhibits a continuous deviation from the ground truth and yields the worst performance due to the absence of position-fixing information. Although \textit{\gls{slam} without \gls{gmf}} significantly reduces the errors, the accumulated errors from the wheel odometry and the gyroscope during the obstructed segment between points L and M still cause a gradual divergence starting from $18$\,seconds. In contrast, \textit{\gls{slam} with \gls{gmf}} halts the error growth at $24.6$\,seconds and subsequently achieves the best performance, benefiting from the \gls{gmf} information generated before $18$\,seconds and reapplied after $24.6$\,seconds. The cumulative frequencies of the agent's absolute position and heading \glspl{rmse} for different $\text{SNR}_{\text{1m}}$ are further shown in the last two subfigures of \Cref{fig:syn_rmse_cdf}, respectively. As observed, \textit{proprioception} exhibits the highest position and heading errors, \textit{\gls{slam} without \gls{gmf}} shows significantly lower errors by incorporating cellular \gls{mf} information, while \textit{\gls{slam} with \gls{gmf}} achieves the lowest errors by further incorporating \gls{gmf} information. As $\text{SNR}_{\text{1m}}$ increases, both position and heading errors decrease for both cases; however, the performance gain is more significant for \textit{\gls{slam} with \gls{gmf}} since a higher \gls{snr} yields more reliable \gls{gmf} information to correct the proprioception input better.     

\subsection{Performance of Real \gls{rf} Measurements}\label{sec:exp_setup}
\subsubsection{Experiment Setup}

\begin{figure} 
\centering
\scalebox{0.9}{\includegraphics[width=\linewidth]{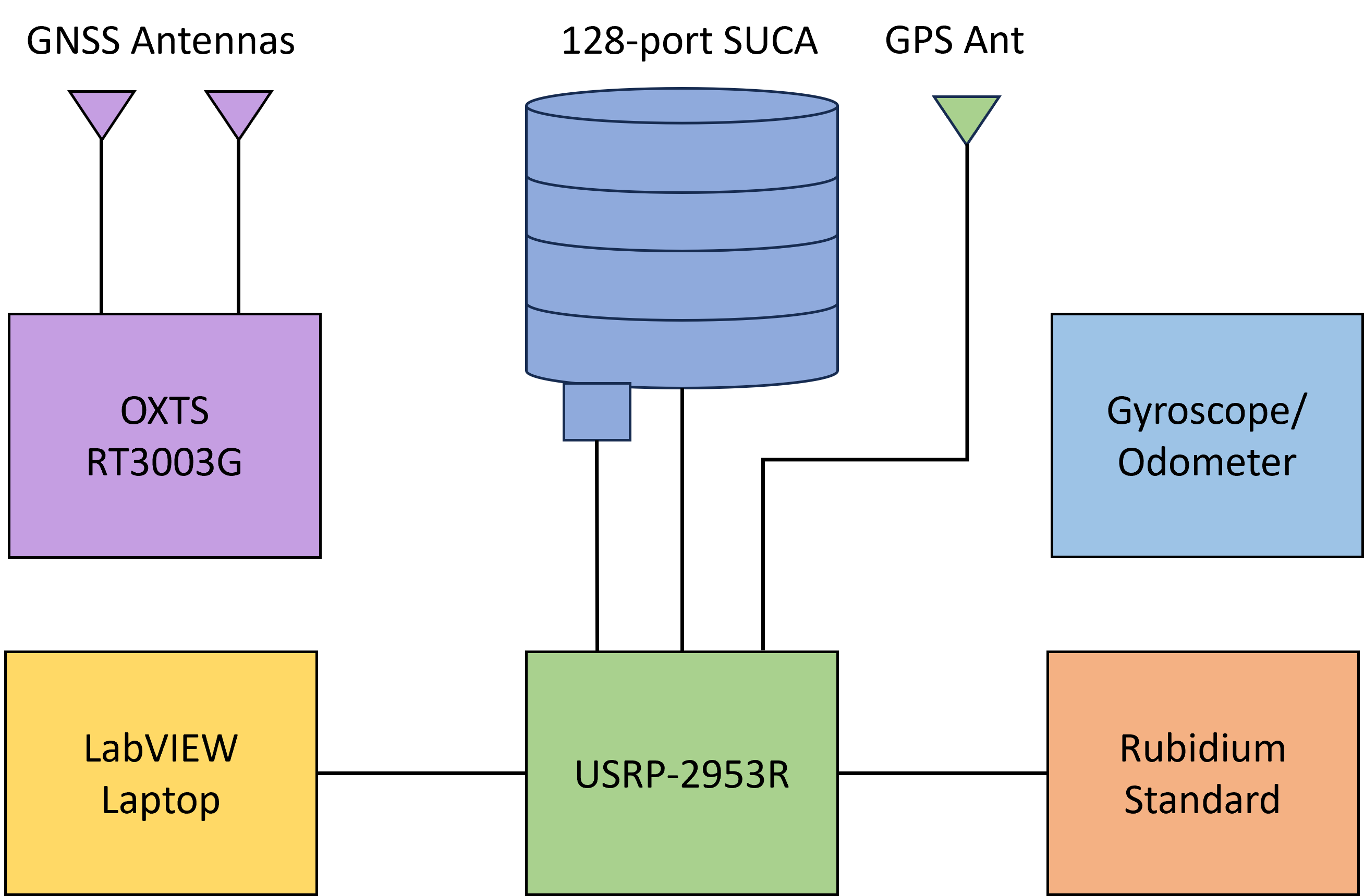}}
\caption{Block diagram of the measurement system, including the ground truth system for pose estimation.}
\label{fig:measure_setup}
\vspace{-5mm}
\end{figure}

A measurement system was developed using \gls{usrp}-2953R from National Instruments \cite{NIUSRP} to control the $128$-port \gls{suca} switch order and log \gls{lte} signals from multiple \glspl{bs}. The block diagram of the measurement system is shown in \Cref{fig:measure_setup}. A GPS disciplined rubidium frequency standard \cite{Rubidium_Link} was used as a stable frequency reference for the \gls{usrp} to minimize clock drift. The antenna array was mounted on the roof of a vehicle, acting as a mobile agent. The $128$ receiving ports were switched in a pseudorandom pattern according to the control signal from the \gls{usrp} with a $0.5$\,ms switching interval, and an additional $11$\,ms was used for the \gls{agc} control. The signals received from multiple commercial \glspl{bs} by the \gls{usrp} were logged on the laptop. The parameters of the cellular system are shown in \Cref{tab:measure_para}. The \gls{usrp} had an internal GPS receiver, and its one pulse per second output was utilized to synchronize the \gls{usrp} itself and the other systems. It also recorded the location information for comparison purposes. The ground truth of the vehicle was generated using an OXTS RT3003G system \cite{oxts}. The vehicle traveled in the urban area of Lund, Sweden at an average speed of about $\qty{1.0}{m/s}$, which was relatively low due to the channel coherence limitations imposed by the switched antenna array system. The longitudinal speed of the vehicle was taken from the wheel odometry, and the yaw rate was retrieved from the gyroscope. Both sensors were mounted in the vehicle.   

\begin{table}[!h]
  \begin{center}
  \renewcommand*{\arraystretch}{1.3}
    \caption{Measurement cellular system information}\label{tab:measure_para}
        \begin{tabular}{|p{4cm}|p{3cm}|}
            \hline
            \textbf{Parameter Name} & \textbf{Value} \\  
            \hline
            Center frequency & $2.66$\,GHz  \\
            \hline
            System bandwidth & $20$\,MHz  \\
            \hline
            BS number & $2$ \\            
            \hline
            Cell IDs of BS A & $375$, $376$, $377$ \\
            \hline
            Cell IDs of BS B & $177$, $178$, $179$\\       
            \hline
            Tx antenna port number & $2$ \\
            \hline
            Rx antenna port number & $128$  \\         
            \hline
            Snapshot interval & $75$\,ms \\
            \hline
            Total snapshot number & $26000$ \\
            \hline            
            Total test time & $32.5$\,minutes \\
            \hline
            Traversed distance & $1750$\,meters \\     
            \hline            
    \end{tabular}
  \end{center}
  \vspace{-5mm}
\end{table}

\begin{figure}[!t]
	\centering
	\hspace*{2mm}\subfloat[\label{subfig:realMeaMap}]
	{\hspace*{-2mm}\includegraphics[width=0.48\textwidth, height=0.42\textwidth]{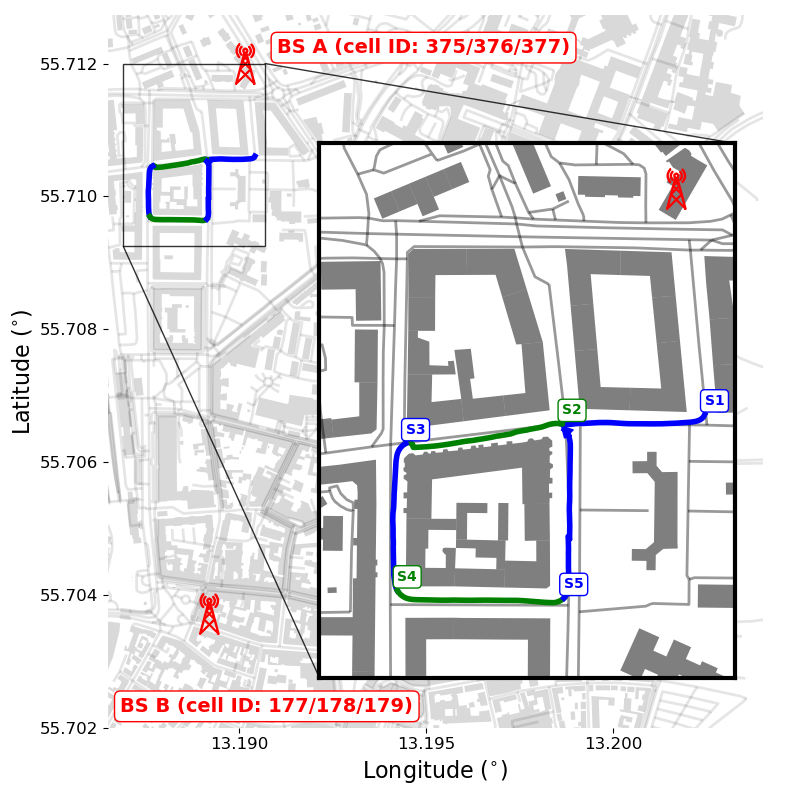}}\\[-2mm]
	\hspace*{5mm}\subfloat[\label{subfig:bs_a_photo}]
	{\hspace*{-4mm}\includegraphics[width=0.22\textwidth, height=0.25\textwidth]{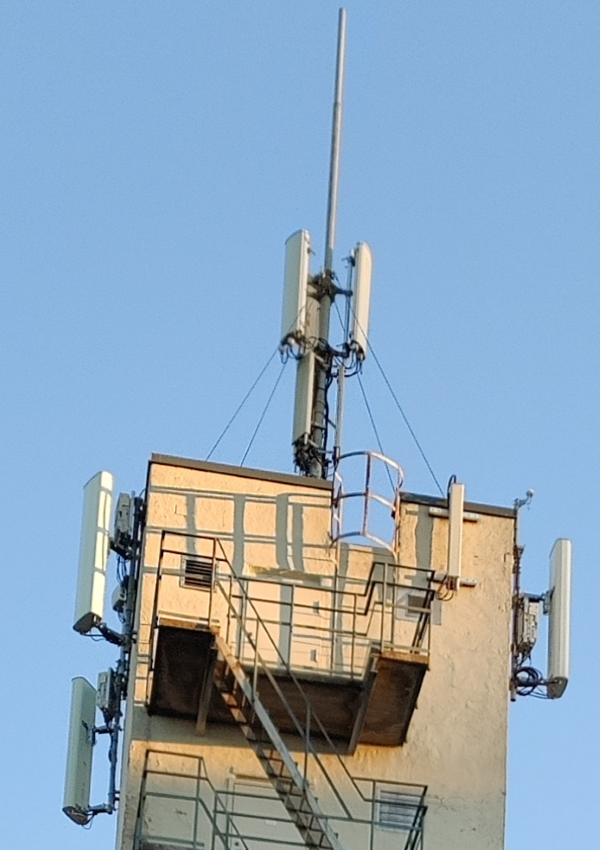}}
	\hspace*{8.5mm}\subfloat[\label{subfig:bs_b_photo}]
	{\hspace*{-4mm}\includegraphics[width=0.22\textwidth, height=0.25\textwidth]{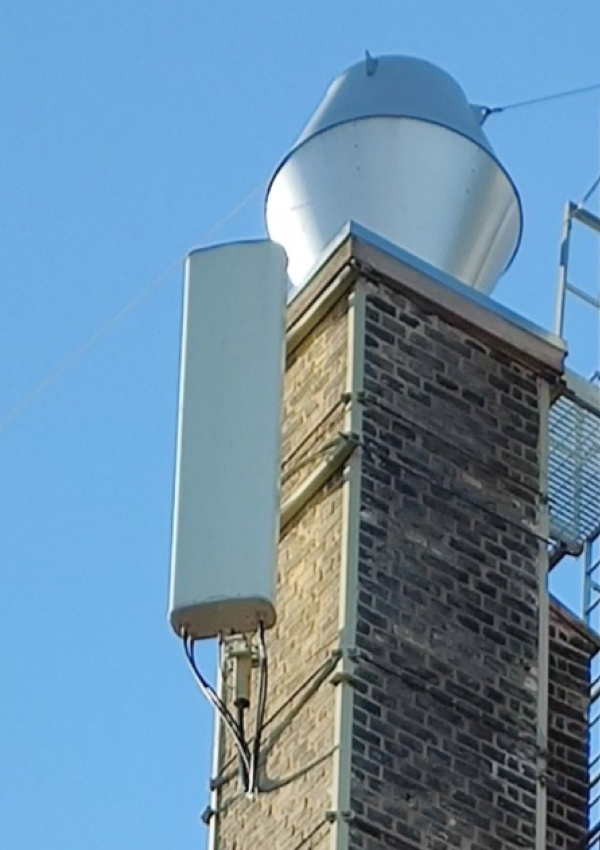}}\\[-1mm]
    \caption{(a) Map of central Lund, Sweden, depicting the true agent trajectory along with the locations of the deployed \glspl{bs}. The agent trajectory is split into $5$ segments from S1 to S5. Photos of \gls{bs} A and \gls{bs} B are shown in (b) and (c).}
	\label{fig:measurement_field}
	\vspace*{-5mm}
\end{figure}

The trajectory of the agent is shown in \Cref{subfig:realMeaMap}, divided into $5$\,segments (S1 to S5) with S2 to S5 constituting a closed loop. The agent starts from S1 and then moves repeatedly along S2 to S5 for four laps. The first and fourth laps are clockwise, and the second and third laps are counterclockwise. Two \glspl{bs} are visible in the measurement field with their photos shown in \Cref{subfig:bs_a_photo,subfig:bs_b_photo}, respectively. \gls{bs} A with cell IDs 375/376/377 is located north of S1, around $160$\,meters from the starting point. \gls{bs} B with cell IDs 177/178/179 is located south of S5 around $800$\,meters. Despite the distance, it is still partially visible to the agent since it is higher than the surrounding buildings. The cell IDs of these two \glspl{bs} are congruent modulo $3$, e.g., $\mod(375,3)\rmv=\rmv\mod(177,3)$, so their \glspl{crs} collide with each other and lead to inter-cell interference. 

For the real \gls{rf} measurement, some parameter settings are different from those of the synthetic \gls{rf} measurement. For instance, the \gls{roi} has a radius $R = 800$\,meters and a height $H=50$\,meters due to the large coverage of the \glspl{bs}. The clutter rate is higher, and the mean value is set to $2$. The noise \glspl{std} of the wheel odometry $\sigma_s$ and the gyroscope $\sigma_{\psi}$ are $\qty{0.05}{m/s}$ and $\qty{0.057}{\degree}$, respectively. The predefined \gls{snr} threshold for adopting a \gls{mpc} as a noisy measurement of \gls{slam} is $19\,\text{dB}$.

\subsubsection{Experimental Results}\label{sec:exp_results}
\begin{figure*}
\centering
\hspace*{1.5mm}\subfloat
{\hspace*{-4.5mm}\includegraphics[width=0.345\textwidth, height=0.31\textwidth]{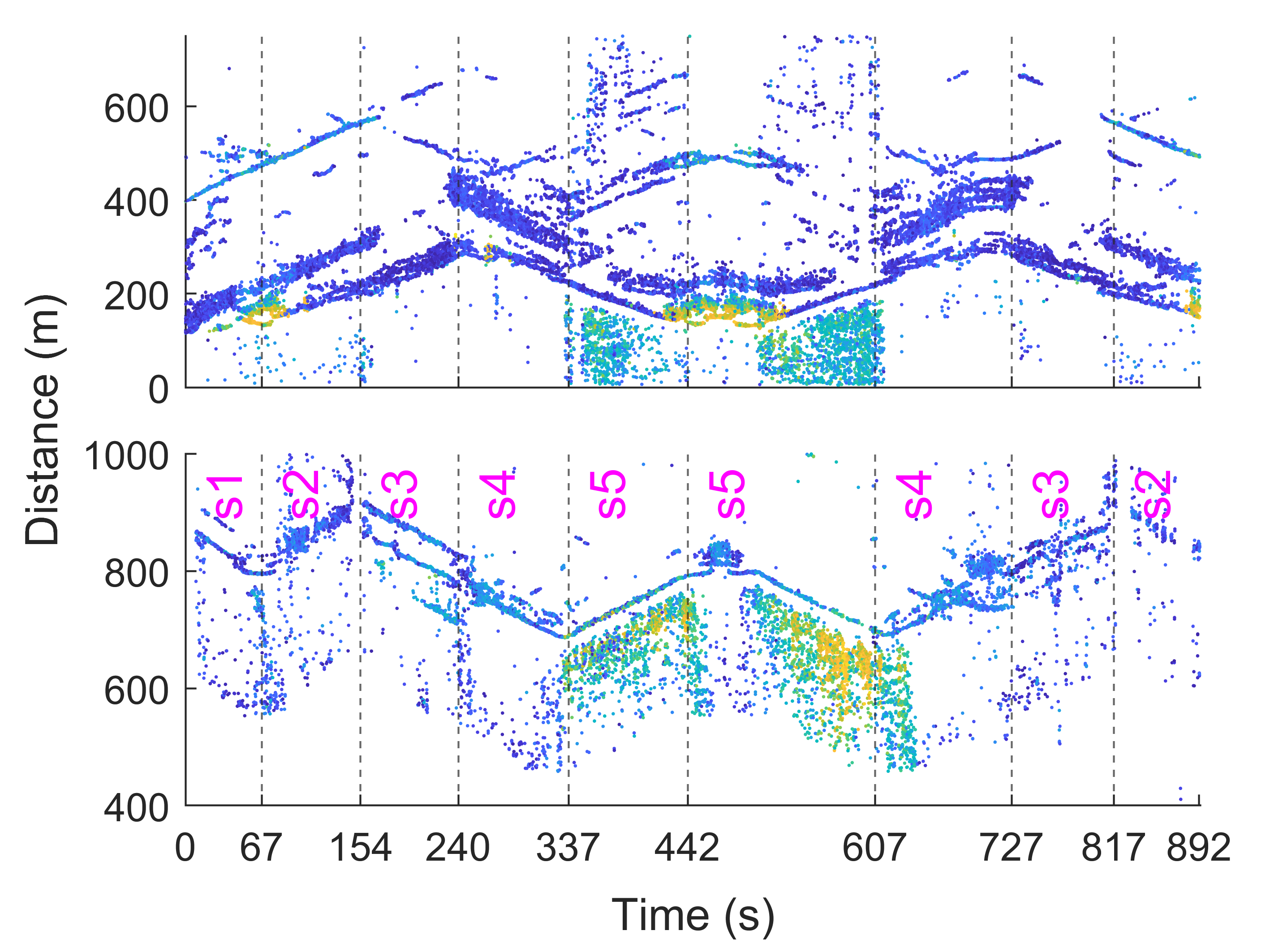}}
\hspace*{1.5mm}\subfloat
{\hspace*{-4.5mm}\includegraphics[width=0.345\textwidth, height=0.31\textwidth]{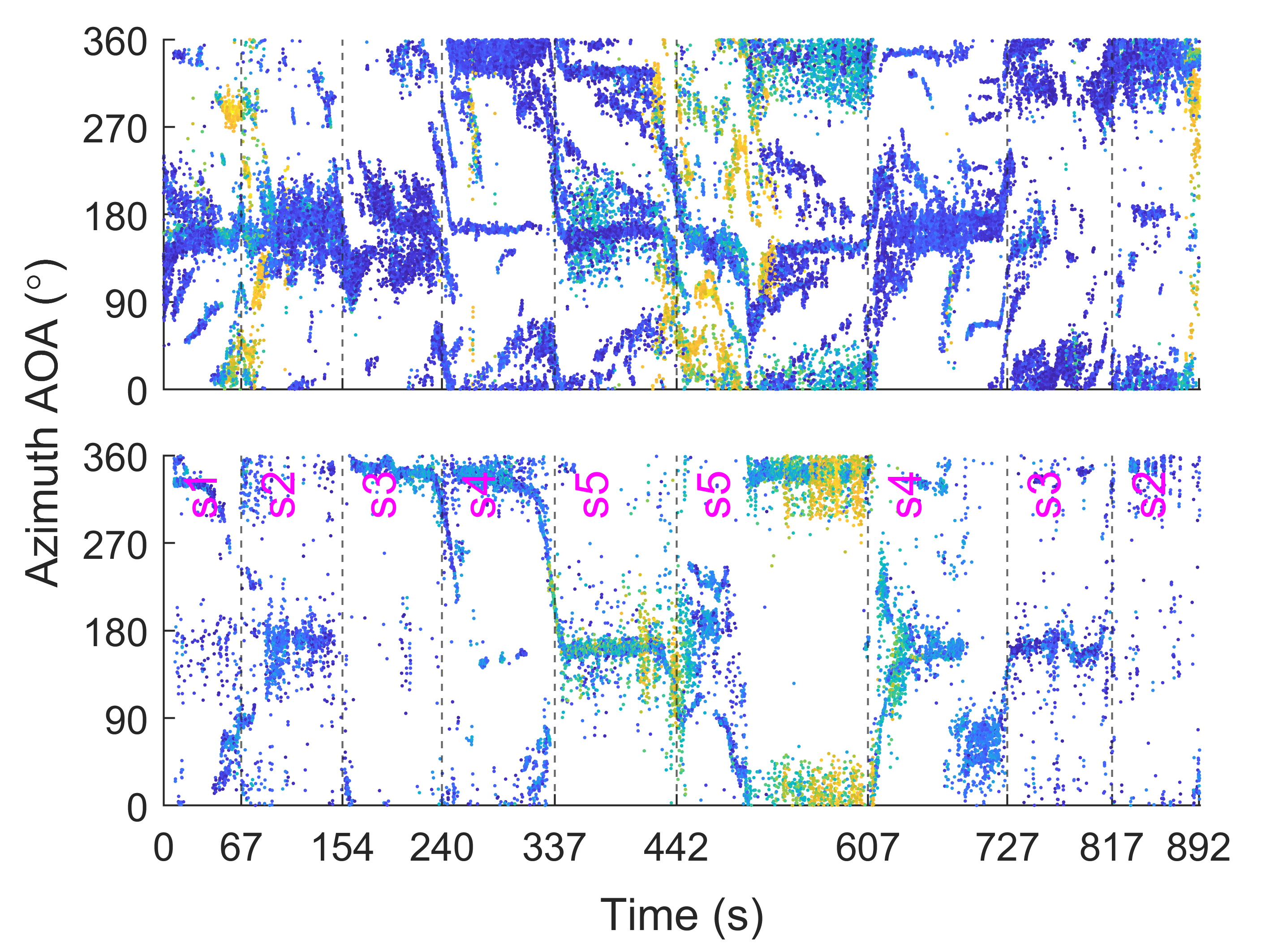}}
\hspace*{-5.5mm}\subfloat
{\hspace*{1.5mm}\includegraphics[width=0.36\textwidth, height=0.31\textwidth]{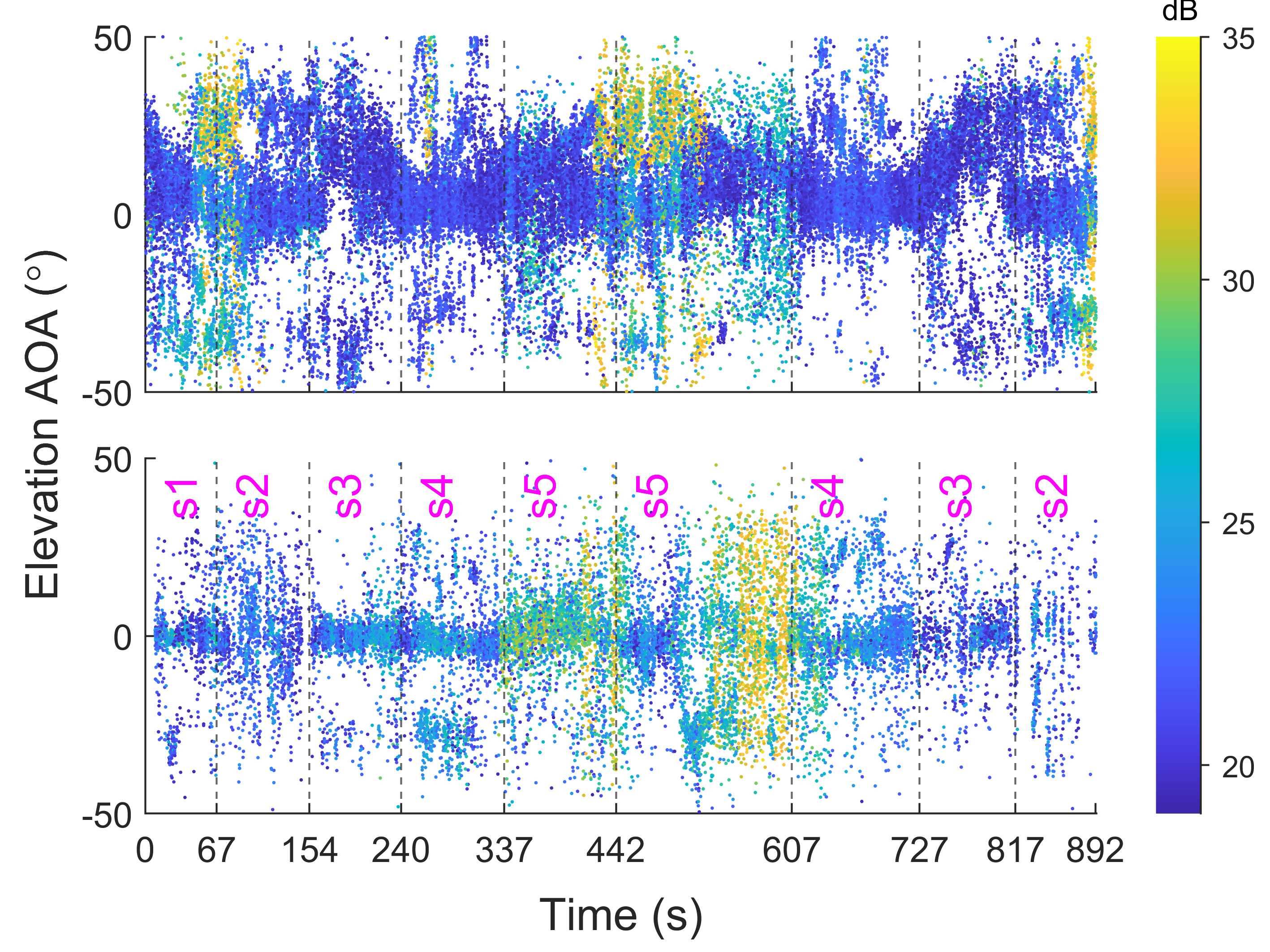}}\\[0mm]
\caption{Results for real \gls{rf} measurements. The RIMAX channel estimator is applied to the received \gls{rf} signals after interference cancellation from cell $376$ of \gls{bs} A (top) and cell $178$ of \gls{bs} B (bottom), providing the \gls{mpc} estimates of distances, azimuth \glspl{aoa}, and elevation \glspl{aoa}. The colormap shows the \gls{snr} estimates of the \glspl{mpc}.}
\label{fig:rimax_estimates}
\vspace{-6mm}
\end{figure*}

The interference cancellation is first applied to cancel the mutual interference between \glspl{crs} from different \glspl{bs}, then the modified RIMAX algorithm \cite{rimax_richter,rimax_ic} is applied to estimate the \gls{mpc} parameters. The estimated distances, azimuth \glspl{aoa} and elevation \glspl{aoa} of \glspl{mpc} associated with cells $376$ and $178$ during the first and second laps are shown in \Cref{fig:rimax_estimates}, respectively, where the colors represent the estimated component \glspl{snr} (i.e., squares of the estimated normalized amplitudes), as indicated in the color map. We can observe that the distance estimates from the same places of two laps exhibit high similarity regardless of the agent's heading. Although the figure displays only the estimated parameters of the first two laps, they closely resemble those of the third and fourth laps, which are not shown to reduce figure clutter.

\begin{figure}
\centerline{\includegraphics[width=0.45\textwidth, height=0.43\textwidth]{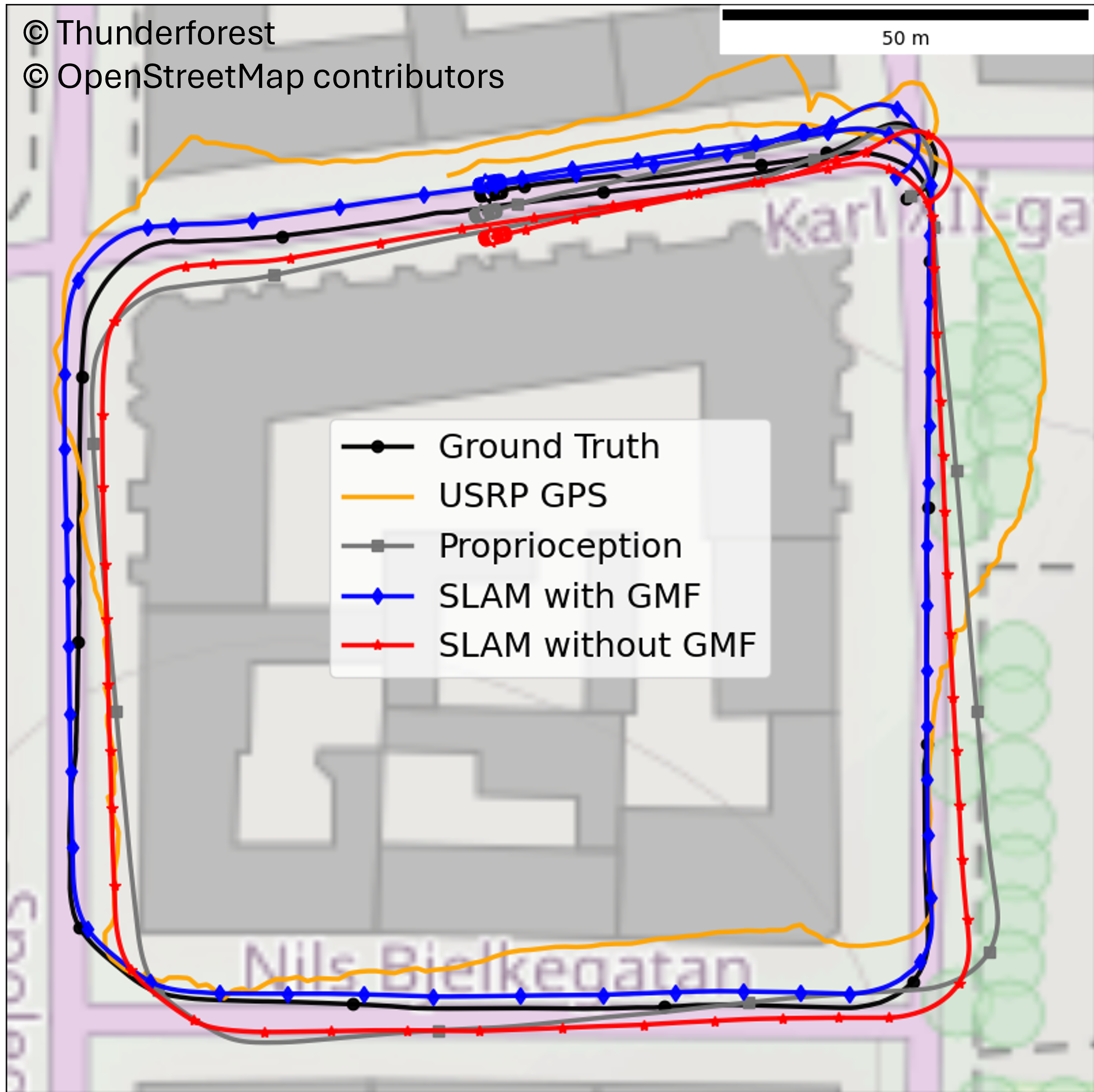}}
\caption{Results for real \gls{rf} measurements. A zoomed-in map of the agent's movement area is presented, showing the true trajectory and the estimated trajectories with various experimental settings.}
\label{fig:osm_ue_traj}
\vspace{-6mm}
\end{figure}

\begin{figure}
    \centering
     \hspace{0mm}\subfloat{\scalebox{1}{\hspace{0mm}\includegraphics{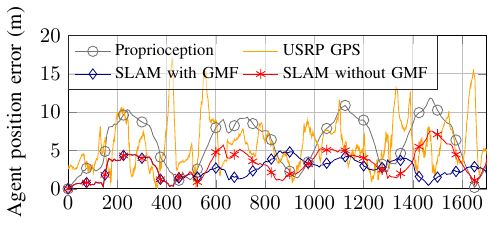}}} \\[-4mm]
    \hspace{0mm}\subfloat{\scalebox{1}{\hspace{0mm}\includegraphics{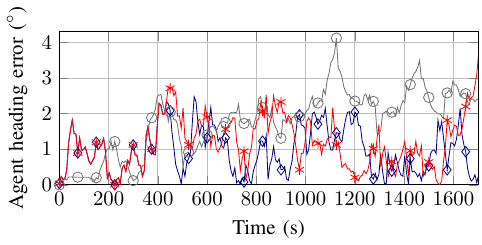}}} \\[0mm]
    \caption{Results for real \gls{rf} measurements. The absolute position errors and heading errors of USRP GPS, proprioception, and \gls{slam} without and with \gls{gmf}.}    \label{fig:abs_dist_heading_error_measurement}
    \vspace{-6mm}
\end{figure}

\begin{figure}[t]
    \centering
    \hspace{0mm}\subfloat{\scalebox{1}{\hspace{1mm}\includegraphics{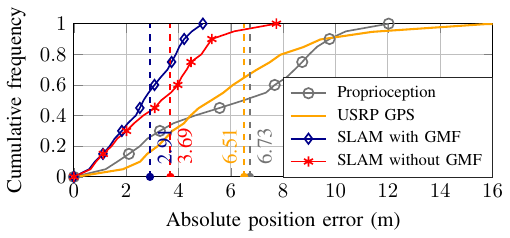}}} \\[-4mm]
    \hspace{0mm}\subfloat{\scalebox{1}{\hspace{0mm}\includegraphics{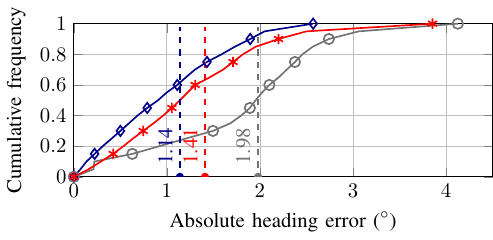}}} \\[0mm]
    \caption{Results for real \gls{rf} measurements. Cumulative frequencies and \glspl{rmse} across different settings for agent position and agent heading.}\label{fig:abs_pos_heading_error_cdf_measurement}	 
    \vspace{-3mm}
\end{figure}

Since the measurement trajectory has minimal altitude variation, we ignore it and assume the height of the agent to be constant during the whole measurement, so we estimate the agent's position in 2D and the \glspl{mf}' positions in 3D (this is just for data processing convenience, and the system model is always valid for 3D position estimation). \Cref{fig:osm_ue_traj} shows the fourth lap's trajectories of the ground truth, USRP GPS, proprioception, and \gls{slam} without and with \gls{gmf}. We can observe that the USRP GPS has the worst performance in this environment due to the limited sky view and heavy multipath, proprioception has the second worst performance, \gls{slam} without \gls{gmf} has the second best performance, and \gls{slam} with \gls{gmf} has the best performance. 
\Cref{fig:abs_dist_heading_error_measurement} shows the absolute position and heading errors of the estimated agent trajectories from \gls{usrp} GPS (as GPS alone provides only point estimates, the corresponding figures omit heading), proprioception, and \gls{slam} without and with \gls{gmf} over time. Their cumulative frequencies and \glspl{rmse} are shown in \Cref{fig:abs_pos_heading_error_cdf_measurement}. The corresponding absolute position and heading \glspl{rmse} are $[6.51, \, 6.73, \,3.69,\, 2.91]$\,meters and $[-, 1.98,\, 1.41,\, 1.14]$\,degrees, respectively. It can be observed that \gls{slam} can provide good localization performance, and the application of \gls{gmf} can further improve the performance.

\begin{figure}
\centerline{\includegraphics[width=\linewidth]{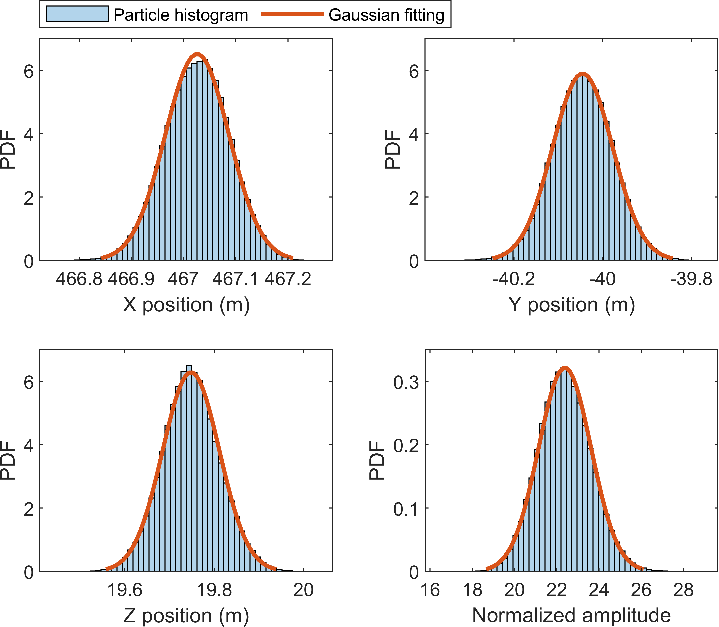}}
\caption{Results for real \gls{rf} measurements. For a converged \gls{gmf} state, the histograms of 3D position and normalized amplitude particles are visualized and individually fitted with a Gaussian distribution.
}
\label{fig:gaussian_fit}
\vspace{-6mm}
\end{figure}

\Cref{fig:gaussian_fit} focuses on a representative \gls{mf} originated from an \gls{mpc} in cell~$376$, which persists from $0$ to $154$~seconds with a distance ranging from $400$ to $600$~meters. The figure shows the histograms of the particles representing the $x$-, $y$-, $z$-coordinates and the normalized amplitudes for this converged \gls{mf}, together with their Gaussian fits. The close match between the histograms and the Gaussian fits validates the assumption in Section \ref{sec:pf_gram} of modeling \gls{gmf} spatial densities as independent Gaussian distributions. The Gaussian approximation not only captures the measurement uncertainty effectively but also enables a compact representation of \glspl{gmf} using only the means and variances of the particle distributions.

\begin{figure*}[t]
\centering
\scalebox{0.75}{\includegraphics[width=\textwidth]{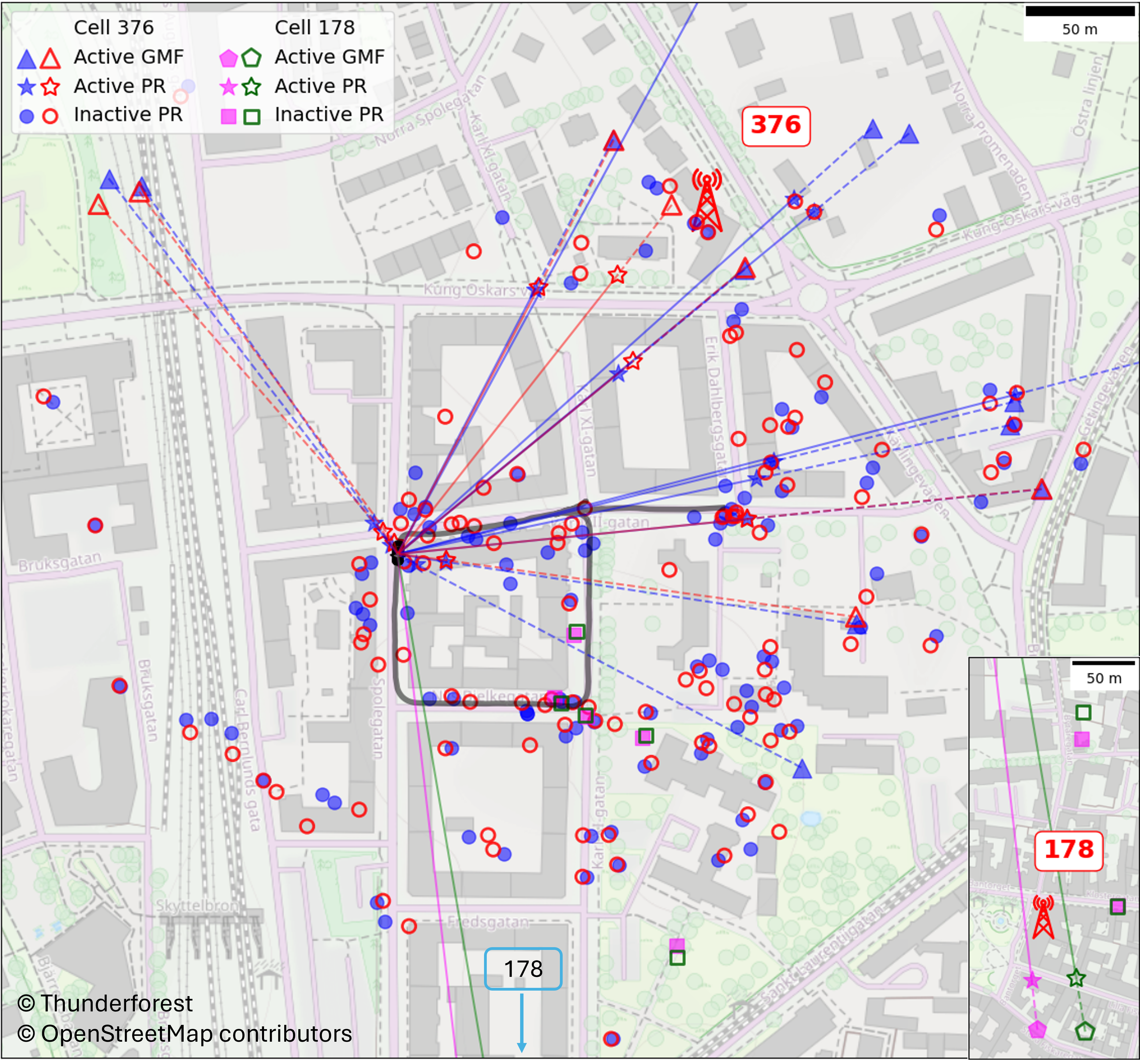}}
\caption{Results for real \gls{rf} measurements in central Lund. Here, lap 1 and lap 2 use filled and empty marks, respectively. All the active \glspl{gmf} at $167$s (lap $1$) and $807$s (lap~$2$) are shown with triangles (cell 376) and pentagons (cell $178$), with their corresponding active \glspl{pr} denoted as stars. The inactive \glspl{pr} from other time instances are shown as circles (cell $376$) and squares (cell $178$). Dashed lines indicate the propagation paths from the \glspl{gmf} to their corresponding \glspl{pr} (mirror paths from the \gls{pa} to the \glspl{pr} are not shown for clarity), and solid lines represent the paths from the \glspl{pr} to the agent. Blue and red lines correspond to cell~$376$, while magenta and green lines correspond to cell~$178$.}\label{fig:gmfs_of_measurement}
\vspace{-4mm}
\end{figure*}

\Cref{fig:gmfs_of_measurement} shows a map of central Lund with all 2D positions of \glspl{pr} associated with cells $376$ and $178$, which are derived from their corresponding \glspl{gmf} at their ending moments during the first two laps. Here, all \glspl{gmf} are assumed first-order reflections of signals from \glspl{bs}. Given the long distance between \gls{bs} B and the agent, a detailed inset of \gls{bs} B is provided in the lower right corner of the figure. This figure illustrates the relative positions of \glspl{pr} to physical objects, such as buildings, in the environment, and also demonstrates the consistency of \glspl{gmf} and \glspl{pr} across two laps, thereby validating the proposed framework. Many \glspl{pr} from these two laps are in close proximity, despite being recorded during different passes. Specifically, results from $167$\,seconds of the first lap and $807$\,seconds of the second lap are presented. The agent positions at these two moments are spatially close ($5.5$\,meters apart in absolute distance) and located in opposite lanes with reversed headings. The positions of the active \glspl{gmf} at these moments (triangles and pentagons of different colors), along with their corresponding positions of \glspl{pr} (stars of different colors), are illustrated in the figure. Dashed lines indicate the paths from \glspl{gmf} to \glspl{pr}, while the solid lines represent the paths from \glspl{pr} to the agent. Many \glspl{gmf} are stably active at both time instances, supporting the validity of the \gls{gmf} concept. However, due to the dynamic nature of the wireless channel, not all \glspl{gmf} observed at $167$\,seconds are active at $807$\,seconds. This temporal variation, which increases the complexity of \glspl{gmf} modeling and utilization, is addressed by the proposed \gls{phd} filter that statistically propagates \glspl{gmf} over time. Furthermore, the \glspl{pr} are mainly located near buildings and align well with the surrounding environment, further supporting the \gls{gmf} concept.

\begin{comment}

\begin{figure*}[t]
\begin{minipage}[c]{0.7\textwidth}
\centering
\scalebox{1}{\includegraphics[width=\textwidth]{images/SLAM_setup_and_results/Measurements/osm_anchors_combined_credit_v2.png}}
\end{minipage}
\hfill
\begin{minipage}[c]{0.28\textwidth} 
\caption{Results for real \gls{rf} measurements in central Lund. Here, lap 1 and lap 2 use filled and empty marks, respectively. All the active \glspl{gmf} at $167$s (lap $1$) and $807$s (lap~$2$) are shown with triangles (cell 376) and pentagons (cell $178$), with their corresponding active \glspl{pr} denoted as stars. The inactive \glspl{pr} from other time instances are shown as circles (cell $376$) and squares (cell $178$). Dashed lines indicate the propagation paths from the \glspl{gmf} to their corresponding \glspl{pr} (mirror paths from the \gls{pa} to the \glspl{pr} are not shown for clarity), and solid lines represent the paths from the \glspl{pr} to the agent. Blue and red lines correspond to cell~$376$, while magenta and green lines correspond to cell~$178$.}\label{fig:gmfs_of_measurement}
\end{minipage}
\vspace{-4mm}
\end{figure*}
\end{comment}

\section{Conclusion}\label{sec_conclusion}
This paper presents a \gls{mpslam} method with a focus on high-accuracy localization in cellular systems within challenging urban environments. The proposed method incorporates \gls{gmf}, and \gls{imu} and wheel odometry information. A \gls{gmf} repository is established with detected \glspl{mf} from early traversals, and a \gls{phd} filter is used to propagate their intensity functions over time in the factor-graph. Comprehensive simulations using synthetic data validate the effectiveness and robustness of the proposed algorithm. Additionally, real-world LTE experiments in central Lund (center frequency $2.66\,\mathrm{GHz}$; system bandwidth $20\,\mathrm{MHz}$) with two base stations whose cell IDs are congruent modulo~3 underline the method's practical viability. In this deployment, colliding \gls{crs} resources from neighboring cells caused severe inter-cell interference. We first canceled mutual interference and then applied a modified RIMAX algorithm to obtain reliable multipath parameters from the cleaned signals. Under these conditions of heavy multipath and limited bandwidth, the proposed algorithm achieved a positioning \gls{rmse} of $2.91\,\mathrm{m}$ and a heading \gls{rmse} of $1.14^\circ$, outperforming the USRP GPS baseline ($6.73\,\mathrm{m}$), proprioception ($6.51\,\mathrm{m}$; $1.98^\circ$), and SLAM without \gls{gmf} ($3.69\,\mathrm{m}$; $1.41^\circ$).  
 
Beyond accuracy, the experiments revealed strong cross-lap repeatability of the multipath geometry and a close match between particle histograms and independent Gaussian fits for GMF positions and normalized amplitudes, lending empirical support to the Gaussian \gls{gmf} model and enabling compact repository storage. Together with the synthetic results, these findings demonstrate that \glspl{gmf} provides informative priors that accelerate redetection after blockages and improve robustness in the presence of inter-cell interference, delivering high-precision localization in real urban deployments. 

Promising directions for future work include exploration of other types of \glspl{mf}, such as point scattering and diffuse scattering, into the proposed \gls{mpslam} framework \cite{LiLeiCaiTuf:ICC2024,Kim2022pmbmslam}, a tight integration of proprioceptive sensor parameters into the \gls{fg} \cite{LeiWieVenWit:Asilomar2024}, or an extension to a hybrid inference framework, e.g., neural enhanced belief propagation \cite{Alexander2024neuralfg, Mingchao2024neuralbpmot}.

\begin{appendices}

\section{Measurement Model}\label{sec:appdx_a}
Given measurement $ \V{z}_{m,n}^{(j)} $, we assume that the conditional \gls{pdf} $ f(\V{z}_{m,n}^{(j)}|\V{x}_{n}, \underline{\V{q}}_{k,n}^{(j)}) $ is conditionally independent across ${z_\mathrm{d}}_{m,n}^{(j)}$, ${z_\mathrm{\varphi}}_{m,n}^{(j)}$, ${z_\mathrm{\vartheta}}_{m,n}^{(j)}$ and ${z_\mathrm{u}}_{m,n}^{(j)}$ given the states ${d}_{k,n}^{(j)}$, ${\varphi}_{k,n}^{(j)}$, ${\vartheta}_{k,n}^{(j)}$ and ${u}_{k,n}^{(j)}$, thus it is factorized as

\begin{align}\label{eq:pdf_factor}
\hspace{-3mm}f(\V{z}_{m,n}^{(j)}\!|\V{x}_{n}, \underline{\V{q}}_{k,n}^{(j)})  = \!f({z_\mathrm{u}}_{m,n}^{(j)} | u_{k,n}^{(j)})  f({z_\mathrm{d}}_{m,n}^{(j)} | \mathrm{d}_{k,n}^{(j)}, u_{k,n}^{(j)}) \notag\\ \times f({z_\mathrm{\varphi}}_{m,n}^{(j)} | \mathrm{\varphi}_{k,n}^{(j)},\! u_{k,n}^{(j)}) f({z_\mathrm{\vartheta}}_{m,n}^{(j)} | \mathrm{\vartheta}_{k,n}^{(j)},\! u_{k,n}^{(j)})\ist.
\end{align}

Assuming Gaussian measurement noise, the individual \glspl{lhf} are given by
\vspace{-1mm}
\begin{align}
    f({z_\mathrm{d}}_{m,n}^{(j)} | d_{k,n}^{(j)}, u_{k,n}^{(j)}) & = C_1 \mathrm{e}^{ \frac{-\big({z_\mathrm{d}}_{m,n}^{(j)} + d_{\text{o}}^{(j)}- d_{k,n}^{(j)}\big)^2}{2\big({\sigma_\mathrm{d}}_{k,n}^{(j)}\big)^2}} \label{eq:pdf_distance} \, \allowdisplaybreaks\\
f({z_\mathrm{\varphi}}_{m,n}^{(j)} | \varphi_{k,n}^{(j)}, u_{k,n}^{(j)}) & = C_2 \mathrm{e}^{\frac{-\big({z_\mathrm{\varphi}}_{m,n}^{(j)}+\varphi_n - \varphi_{k,n}^{(j)}\big)^2}{2\big({\sigma_\mathrm{\varphi}}_{k,n}^{(j)}\big)^2}} \label{eq:pdf_azu} \,   \\
f({z_\mathrm{\vartheta}}_{m,n}^{(j)} | \vartheta_{k,n}^{(j)}, u_{k,n}^{(j)}) & = C_3 \mathrm{e}^{\frac{-\big({z_\mathrm{\vartheta}}_{m,n}^{(j)} + \theta_n - \vartheta_{k,n}^{(j)}\big)^2}{2\big({\sigma_\mathrm{\vartheta}}_{k,n}^{(j)}\big)^2}}  \label{eq:pdf_elv} 
    \\[-7mm]\nn
\end{align}
where $C_1 = 1/(\sqrt{2\pi}{\sigma_\mathrm{d}}_{k,n}^{(j)}), C_2 = 1/(\sqrt{2\pi}{\sigma_\mathrm{\varphi}}_{k,n}^{(j)})$ and $C_3 = 1/(\sqrt{2\pi}{\sigma_\mathrm{\vartheta}}_{k,n}^{(j)})$. 
The variances depend on $u_{k,n}^{(j)}$ and are determined based on the Fisher information as in \cite{LiLeiCaiTuf:ICC2024}. 

The \gls{lhf} $ f({z_\mathrm{u}}_{m,n}^{(j)} | u_{k,n}^{(j)}) $ of the normalized amplitude measurement $ {z_\mathrm{u}}_{m,n}^{(j)} $ is modeled by a truncated Rician \gls{pdf} \cite[Ch.\,1.6.7]{BarShalom2011AlgorithmHandbook}, i.e.,
\vspace{-1mm}
\begin{align}
    &f({z_\mathrm{u}}_{m,n} ^{(j)}| u_{k,n}^{(j)}) \nn\\[-1mm]
    &\hspace{8mm}= \frac{  \frac{{z_\mathrm{u}}_{m,n}^{(j)}}{({\sigma_\mathrm{u}}_{k,n}^{(j)})^2} \mathrm{e}^{\big(\frac{ -\big(({z_\mathrm{u}}_{m,n}^{(j)})^2 + (u_{k,n}^{(j)})^2\big)}{2({\sigma_\mathrm{u}}_{k,n}^{(j)})^2} \big)} \mathrm{I}_{0}(\frac{{z_\mathrm{u}}_{m,n}^{(j)} u_{k,n}^{(j)}}{({\sigma_\mathrm{u}}_{k,n}^{(j)})^2}) } { P_{\mathrm{d}}(u_{k,n}^{(j)}) }
	\label{eq:pdf_normAmplitude}\\[-7mm]\nn
\end{align} 
for $ {z_\mathrm{u}}_{m,n}^{(j)} > \sqrt{u_{\mathrm{de}}} $, where $ ({\sigma_\mathrm{u}}_{k,n}^{(j)})^2 = \frac{1}{2} + \frac{1}{4N_{\mathrm{f}}N_{\mathrm{a}}}(u_{k,n}^{(j)})^2 $ \cite{xuhong2022twc}, $ \mathrm{I}_{0}(\cdot) $ represents the $ 0 $th-order modified first-kind Bessel function, and $u_{\mathrm{de}}$ is the detection threshold of the channel estimator. The detection probability $P_{\mathrm{d}}(u_{k,n}^{(j)})$ is modeled by a Rician \gls{cdf} \cite{erik_icc_2019, xuhong2022twc},
\vspace{-1mm}
\begin{align} \label{eq:pd_u}
P_{\mathrm{d}}(u_{k,n}^{(j)}) = Q_{1}\big(u_{k,n}^{(j)}/{\sigma_\mathrm{u}}_{k,n}^{(j)}, \sqrt{u_{\mathrm{de}}}/{\sigma_\mathrm{u}}_{k,n}^{(j)}\big) \\[-7mm]\nn
\end{align} 
where $ Q_{1}(\boldsymbol{\cdot},\boldsymbol{\cdot}) $ denotes the Marcum Q-function \cite[Ch.\, 1.6.7]{BarShalom2011AlgorithmHandbook}. 

False alarm measurements are modeled by a Poisson point process with mean $ {\rv{\mu}^{(j)}_{\mathrm{fa}\ist {n}}} $ and \gls{pdf} $ f_{\mathrm{fa}}(\V{z}_{m,n}^{(j)}) $, which is factorized as 
\begin{align}\label{eq:pdf_fa}
\resizebox{.88\hsize}{!}{$%
f_{\mathrm{fa}}(\V{z}_{m,n}^{(j)}) = f_{\mathrm{fa}}^{\mathrm{d}}({z_\mathrm{d}}_{m,n}^{(j)}) f_{\mathrm{fa}}^{\mathrm{\varphi}}({z_\varphi}_{m,n}^{(j)})f_{\mathrm{fa}}^{\mathrm{\vartheta}}({z_\vartheta}_{m,n}^{(j)})f_{\mathrm{fa}}^{\mathrm{u}}({z_\mathrm{u}}_{m,n}^{(j)})$%
}\\[-7mm]\nn
\end{align} 
where $ f_{\mathrm{fa}}^{\mathrm{d}}({z_\mathrm{d}}_{m,n}^{(j)})\rmv\rmv=\rmv\rmv 1/d_{\mathrm{max}}^{(j)} $, $ f_{\mathrm{fa}}^{\mathrm{\varphi}}({z_\mathrm{\varphi}}_{m,n}^{(j)})\rmv\rmv =\rmv\rmv 1/(2\pi)$ and $ f_{\mathrm{fa}}^{\mathrm{\vartheta}}({z_\mathrm{\vartheta}}_{m,n}^{(j)})\rmv\rmv =\rmv\rmv 1/\pi $ are assumed to be uniform on $[0,\ist d_{\mathrm{max}}^{(j)}]$, $[0,\ist 2\pi)$ and $[-\pi/2,\ist \pi/2]$, respectively. The false alarm \gls{pdf} of the normalized amplitude is a Rayleigh \gls{pdf} given as $ f_{\mathrm{fa}}^{\mathrm{u}} ({z_\mathrm{u}}_{m,n}^{(j)})\rmv\rmv=\rmv\rmv 2{z_\mathrm{u}}_{m,n}^{(j)} e^{-({z_\mathrm{u}}_{m,n}^{(j)})^2}/p_{\mathrm{fa}}$ for ${z_\mathrm{u}}_{m,n}^{(j)}\rmv\rmv >  \sqrt{u_{\mathrm{de}}}$, and $ p_{\mathrm{fa}} = e^{-u_{\mathrm{de}}} $ denotes the false alarm probability \cite{BarShalom2011AlgorithmHandbook}.

\section{Update Agent State with New \gls{pf}}\label{sec:appdx_b}
In the absence of information on the environmental geometry, previous works \cite{Erik2019SLAM_TWC, LeiVenTeaMey:TSP2023} have generally assumed that the prior \glspl{pdf} of the new \glspl{pf} $f_{\mathrm{u},n}(\bar{\V{q}}^{(j)}_{m,n})$ are uniformly distributed over the \gls{roi}, and the newly initialized \glspl{pf} do not immediately contribute to the update of the agent state, that is, only messages from legacy \glspl{pf} are used at time $n$. In this work, the \gls{phd} filters provide the informative prior \glspl{pdf} $f_{\mathrm{u},n}(\bar{\V{q}}^{(j)}_{m,n})$ for the new \glspl{pf}, therefore they are also exploited for agent update, which is similar to \cite{xuhong2022twc, Kim2024TSP}. The detailed derivation is shown below. 

For the agent, the messages $\rho_{k}^{(j)}(\V{x}_{n})$ passed from the legacy \gls{pf}-related factor nodes $g\big( \V{x}_{n}, \underline{\V{q}}^{(j)}_{k,n}, \underline{r}^{(j)}_{k,n}, \underline{a}^{(j)}_{k,n}; \V{z}^{(j)}_{n} \big)$ to the variable node $\V{x}_{n}$ are calculated in line with \cite[Eq.\,32]{Erik2019SLAM_TWC} 
\begin{align}	
\rho_{k}^{(j)}(\V{x}_{n}) & = \rmv\rmv\rmv\rmv\rmv\rmv\rmv\rmv \sum_{\underline{a}^{(j)}_{k,n}=0}^{M_{n}^{(j)}} 
\rmv\rmv\rmv\rmv\rmv\rmv \eta(\underline{a}^{(j)}_{k,n}) \rmv\rmv\rmv\rmv\rmv\rmv \sum_{\underline{r}^{(j)}_{k,n} \rmv\rmv \in \{0,1\} }\rmv\rmv\rmv\rmv\rmv\rmv\rmv\rmv\rmv\rmv\rmv\rmv \int g\big( \V{x}_{n}, \underline{\V{q}}^{(j)}_{k,n}, \underline{r}^{(j)}_{k,n}, \underline{a}^{(j)}_{k,n}; \V{z}^{(j)}_{n} \big)  \nonumber \\[1mm]
& \hspace*{2mm} \times \alpha_{k}(\underline{\V{q}}^{(j)}_{k,n}, \underline{r}^{(j)}_{k,n}) \mathrm{d} \underline{\V{q}}^{(j)}_{k,n} \nonumber \\[1mm]
& = \rmv\rmv \sum_{\underline{a}^{(j)}_{k,n}=0}^{M_{n}^{(j)}} 
 \eta(\underline{a}^{(j)}_{k,n}) \int g\big( \V{x}_{n}, \underline{\V{q}}^{(j)}_{k,n}, 1, \underline{a}^{(j)}_{k,n}; \V{z}^{(j)}_{n} \big)  \nonumber \\[1mm]
& \hspace*{2mm} \times \alpha_{k}(\underline{\V{q}}^{(j)}_{k,n},1) \mathrm{d} \underline{\V{q}}^{(j)}_{k,n} + \eta(\underline{a}^{(j)}_{k,n}=0) \alpha_{k,n}
\label{eq:g2agent} \vspace*{0mm}
\end{align}
where $\alpha_{k}(\underline{\V{q}}^{(j)}_{k,n}, \underline{r}^{(j)}_{k,n})$ denotes the prediction messages for the legacy \glspl{pf}, and $\alpha_{k,n} = \int \alpha_{k}(\underline{\V{q}}^{(j)}_{k,n},0) \mathrm{d} \underline{\V{q}}^{(j)}_{k,n}$. The messages $\kappa_{m}^{(j)}(\V{x}_{n})$ passed from the new \gls{pf}-related factor nodes $h\big( \V{x}_{n}, \bar{\V{q}}^{(j)}_{m,n}, \bar{r}^{(j)}_{m,n}, \overline{a}^{(j)}_{m,n}; \V{z}^{(j)}_{n} \big)$ to the variable node $\V{x}_{n}$ are calculated by 
\vspace{-1mm}
\begin{align}	
\kappa_{m}^{(j)}(\V{x}_{n}) & = \varsigma(\overline{a}^{(j)}_{m,n}=0)\int h\big( \V{x}_{n}, \bar{\V{q}}^{(j)}_{m,n},1, 0; \V{z}^{(j)}_{n} \big) \mathrm{d} \bar{\V{q}}^{(j)}_{m,n} \nonumber \\
& \hspace*{2mm} + \sum_{\overline{a}^{(j)}_{m,n}=0}^{K_{n-1}^{(j)}} \varsigma(\overline{a}^{(j)}_{m,n})
\label{eq:h2agent}\\[-7mm]\nn
\end{align}
where the probabilistic \gls{da} messages $\eta(\underline{a}^{(j)}_{k,n})$ and $\varsigma(\overline{a}^{(j)}_{m,n})$ are obtained with an efficient loopy \gls{bp}-based algorithm as shown in \cite{Florian2017TSP, Erik2019SLAM_TWC}. With the messages above, the belief $q(\V{x}_{n})$ approximating the marginal posterior \gls{pdf} $ f(\V{x}_{n}|\V{z}_{1:n} )$ is obtained, up to a normalization factor, as
\vspace{-2mm}
\begin{align}	
q(\V{x}_{n}) & \propto \alpha(\V{x}_{n}) \prod_{j=1}^{J} \prod_{k=1}^{K_{n-1}^{(j)}} \rho_{k}^{(j)}(\V{x}_{n}) \prod_{m=1}^{M_{n}^{(j)}} \kappa_{m}^{(j)}(\V{x}_{n})\, .
\label{eq:MMSE_x} \vspace*{0mm}
\end{align}

\begin{figure}[t]
	\centering
	  \scalebox{0.53}{\hspace{0mm}\includegraphics{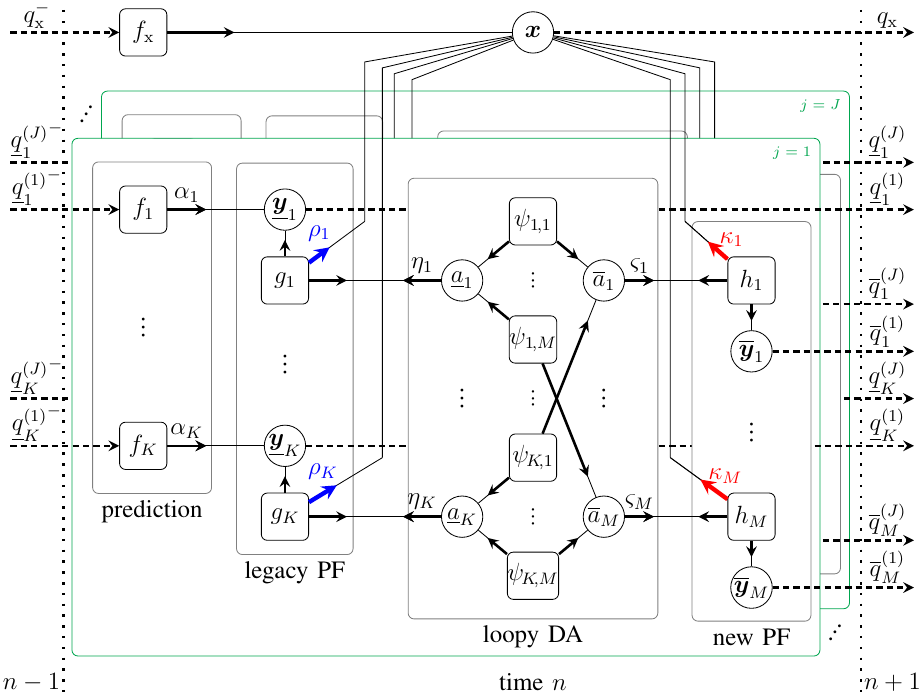}}
	\caption{Factor graph representation of the factorized joint posterior \gls{pdf} (\ref{eq:joint_pdf}). For simplicity, the following short notations are used: $ K \triangleq K_{n-1}^{(j)} $, $ M \triangleq M_{n}^{(j)} $; $\alpha_{k} \triangleq \alpha_{k}(\underline{\V{q}}^{(j)}_{k,n}, \underline{r}^{(j)}_{k,n})$; \emph{factor nodes}: $ f_{\mathrm{x}} \triangleq f(\V{x}_{n}|\V{x}_{n-1}) $, $ f_{k} \triangleq f(\underline{\V{y}}^{(j)}_{k,n}|\V{y}^{(j)}_{k,n-1}) $, $ g_{k} \triangleq g\big( \V{x}_{n}, \underline{\V{q}}^{(j)}_{k,n}, \underline{r}^{(j)}_{k,n}, \underline{a}^{(j)}_{k,n}; \V{z}^{(j)}_{n} \big) $, $ h_{m} \triangleq h\big( \V{x}_{n}, \bar{\V{q}}^{(j)}_{m,n}, \bar{r}^{(j)}_{m,n}, \overline{a}^{(j)}_{m,n}; \V{z}^{(j)}_{n} \big) $, $ \psi_{k,m} \triangleq \psi(\underline{a}^{(j)}_{k,n},\overline{a}^{(j)}_{m,n}) $; \emph{loopy DA}: $\eta_{k} = \eta(\underline{a}^{(j)}_{k,n})$, $\varsigma_{m} = \varsigma(\overline{a}^{(j)}_{m,n})$ ; \emph{measurement update for agent}: $ \rho_{k} \triangleq \rho_{k}^{(j)}(\V{x}_{n}) $, $ \kappa_{m} \triangleq \kappa_{m}^{(j)}(\V{x}_{n}) $; belief calculation: $ q^{-}_{\mathrm{x}} \triangleq q(\V{x}_{n-1}) $, $ q_{\mathrm{x}} \triangleq q(\V{x}_{n}) $, $ {\underline{q}^{(j)}_{k}}^{-} \triangleq \underline{q}(\underline{\V{q}}^{(j)}_{k,n-1}, \underline{r}^{(j)}_{k,n-1}) $, $ \underline{q}^{(j)}_{k} \triangleq \underline{q}(\underline{\V{q}}^{(j)}_{k,n}, \underline{r}^{(j)}_{k,n}) $, $ \overline{q}^{(j)}_{m} \triangleq \overline{q}(\bar{\V{q}}^{(j)}_{m,n}, \bar{r}^{(j)}_{m,n}) $. The other nodes and messages represented by the black arrow lines are formulated in line with \cite{Erik2019SLAM_TWC}.}	 
	\label{fig:factorGraph}
\end{figure}

\end{appendices}

\section*{Acknowledgment} \label{sec:ack}
The authors would like to thank Martin Nilsson at Lund University for his help in setting up the measurement system.

\renewcommand{\baselinestretch}{0.97} 
\bibliographystyle{IEEEtran}
\bibliography{IEEEabrv,liberry}

\end{document}